\documentclass[11pt]{article}
\usepackage{epsfig}

\usepackage{listings}
\usepackage{xcolor}

\usepackage{dcolumn}
\usepackage{bm}
\usepackage{adjustbox}
\usepackage{booktabs}
\usepackage{graphicx}
\usepackage[textsize=tiny]{todonotes}
\usepackage{subcaption}
\usepackage{here}
\usepackage{amssymb,amsmath}
\usepackage{multirow}
\usepackage{cite,color,url}
\usepackage{tabu}
\usepackage{array}
\usepackage{booktabs}
\usepackage{slashed}
\usepackage{xspace}
\usepackage[colorlinks=true,urlcolor=blue,anchorcolor=blue
,citecolor=blue,filecolor=blue,linkcolor=blue,menucolor=blue
,linktocpage=true,pdfproducer=medialab,pdfa=true]{hyperref}
\usepackage{colordvi}
\usepackage{epsfig,psfrag,rotating,soul}
\def\beqa{\begin{eqnarray}}
\def\eeqa{\end{eqnarray}}

\newcommand{\MeV}{{\ensuremath\,\rm MeV}\xspace}
\newcommand{\GeV}{{\ensuremath\,\rm GeV}\xspace}
\newcommand{\TeV}{{\ensuremath\,\rm TeV}\xspace}
\newcommand{\pb}{{\ensuremath\,\rm pb}\xspace}
\newcommand{\fb}{{\ensuremath\,\rm fb}\xspace}

\newcommand{\eqn}{equation}
\evensidemargin \oddsidemargin
\allowdisplaybreaks
\renewcommand{\arraystretch}{1.5}
\psfrag{lk}{$l_k$}
\psfrag{lm}{$l_m$}
\psfrag{blk}{$\bar {l}_k$}
\psfrag{blm}{$\bar {l}_m$}
\def\thefootnote{\fnsymbol{footnote}}

\let\OLDthebibliography\thebibliography
\renewcommand\thebibliography[1]{
\OLDthebibliography{#1}
\setlength{\parskip}{0pt}
\setlength{\itemsep}{0pt plus 0.3ex}}
\usepackage{makecell}

\newcommand{\lb}{\left(}
\newcommand{\rb}{\right)}
\begin{document}
\rightline{RBI-ThPhys-2026-28, COMETA-2026-31, CQUeST-2026-0787}

\thispagestyle{empty}
\vspace{5mm}
\begin{center}
\begin{Large}
\textbf{\textsc{Search for Light Scalars in the Two Real Singlet Model\\ \vspace{2mm} at the LHC}}
\end{Large}

\vspace{1cm}
{
Aman Desai,$^{1}$%
\footnote{\tt
\href{mailto:aman.desai@adelaide.edu.au}{aman.desai@adelaide.edu.au}}
Kristin Lohwasser,$^{2}$%
\footnote{\tt
	\href{mailto:kristin.lohwasser@cern.ch}{kristin.lohwasser@cern.ch}}
Mohamed Ouchemhou,$^{3}$%
\footnote{\tt
\href{mailto:Mohamed.Ouchemhou@irb.hr}{Mohamed.Ouchemhou@irb.hr}}
}
Tania Robens,$^{3}$%
\footnote{\tt
\href{mailto:Tania.Natalie.Robens@irb.hr}{trobens@irb.hr}}
and 

Prasenjit Sanyal,$^{4, 5}$%
\footnote{\tt
\href{mailto:prasenjit.sanyal01@gmail.com}{prasenjit.sanyal01@gmail.com}}
\vspace*{.7cm}

{\sl
	$^{1}$ Department of Physics, Adelaide University, North Terrace, Adelaide, SA 5005, Australia}\\

{\sl 
    $^{2}$ School of Mathematical and Physical Sciences, University of Sheffield, Hicks Building, Hounsfield Road, Sheffield, S3
7RH, UK}\\
{\sl
$^3$ Ruder Boskovic Institute, Bijenicka cesta 54, 10000 Zagreb, Croatia}

{\sl
	$^4$ Center for Quantum Spacetime, Sogang University, Seoul 121-742, South Korea}

{\sl
	$^5$ Department of Physics, Sogang University, Seoul 121-742, South Korea}
\end{center}

\vspace*{0.1cm}
\begin{abstract}

We investigate exotic scalar decays $h_2 \to h_1 h_1$ in the Two Real Singlet Model (TRSM), focusing on light CP-even scalars with $20 \le M_{1} \le 60$~GeV and $20 \le M_{2} \le 120$~GeV. Unlike previous studies focused on gluon--gluon fusion or vector-boson fusion production, we explore associated production with electroweak gauge bosons, $pp \to Vh_2$ {\sl($V = W,Z$)}, as a complementary and experimentally clean probe of light scalar cascades in the TRSM. The subsequent decay $h_1 \to b\bar{b}$ leads to final states containing four $b$-jets accompanied by a charged lepton and missing transverse energy. We evaluate the sensitivity of the $4b+\ell\nu_\ell$ channel at $\sqrt{s}=13.6$~\TeV for the LHC Run~3 and the HL-LHC. 

For an integrated luminosity of $300~\mathrm{fb}^{-1}$, the $W^+h_2$ channel provides the strongest sensitivity, reaching a significance of approximately $4\sigma$, while the $W^-h_2$ channel reaches approximately $3\sigma$. At the HL-LHC with $3000~\mathrm{fb}^{-1}$, the corresponding significances increase to approximately $12\sigma$ and $8\sigma$, respectively, highlighting the strong discovery potential of associated light-scalar production in the TRSM.
\end{abstract}

\def\thefootnote{\arabic{footnote}}
\setcounter{page}{0}
\setcounter{footnote}{0}
\newpage
\tableofcontents

\newpage
\section{Introduction}
Since the discovery of the Higgs boson by the ATLAS and CMS Collaborations in 2012 \cite{ATLAS:2012yve,CMS:2012qbp}, measurements at the Large Hadron Collider (LHC) have shown excellent agreement with the predictions of the Standard Model (SM) of particle physics \cite{ATLAS:2022vkf, CMS:2022dwd}. Nevertheless, the Higgs sector remains a key area for explaining new puzzles and searches for physics beyond the SM (BSM). 

The ongoing search for physics BSM has therefore motivated studies that go beyond SM decay topologies, 
focusing instead on BSM signatures $h_i\to h_jh_k$. One particularly compelling avenue involves the existence of additional scalar bosons beyond the observed 125~GeV Higgs particle. Light scalar states, with masses below that of the SM Higgs boson, arise naturally in a variety of theoretical frameworks~ {\sl(see e.g. \cite{Curtin:2013fra,Cepeda:2021rql, Robens:2025nev})} and can give rise to rich collider phenomenology, including exotic decay modes. Such particles may play a crucial role in addressing fundamental open questions, for instance by providing portals to hidden or dark sectors, or by enabling a strong first-order electroweak phase transition required for successful baryogenesis.

Among the simplest and most well-motivated extensions are models in which additional gauge-singlet scalar fields are introduced~\cite{Patt:2006fw, OConnell:2006rsp, Barger:2007im}. Such singlet extensions preserve the SM gauge structure while giving rise to rich phenomenological consequences through Higgs–portal interactions~\cite{Chen:2014ask, Robens:2015gla, Robens:2016xkb}. In particular, models with more than one singlet scalar can significantly enlarge the scalar sector dynamics, allowing for non-trivial vacuum structures,  and distinctive collider signatures, while remaining consistent with current experimental constraints~\cite{Robens:2019kga}. Our focus is on a specific extension of the SM by two real scalar singlets (TRSM)~\cite{Robens:2019kga,Robens:2022nnw,Robens:2025tew}.

Recent searches for the exotic decay $h \to aa$, where $a$ is a light scalar or pseudoscalar, have been performed by the ATLAS and CMS collaborations in multiple final states. These include fully hadronic final states such as $4b$~\cite{ATLAS:2025rfm,ATLAS:2018pvw,CMS:2022xxa}, as well as mixed channels like $\mu\mu bb$ and $\tau\tau bb$~\cite{CMS:2024uru}, complementary
searches target leptonic and photonic signatures, including $bb\tau\tau$~\cite{ATLAS:2024vpj}, $\gamma\gamma\tau\tau$~\cite{CMS:2024uru}, and clean photonic channels such as $4\gamma$~\cite{CMS:2026knm,ATLAS:2023ian}. In all cases, no significant excess over the SM expectation is observed, and upper limits are set on $\mathrm{BR}(h \to aa)$ across a wide range of $m_a$. Unlike previous studies, we explore complementary searches using associated production with a vector boson in the TRSM~\cite{Desai:2026fvt}. 

In this paper, we investigate the production of a light scalar boson $h_2$ in association with $W$ and $Z$ bosons at the LHC, where $h_2$ subsequently decays into a pair of lighter scalars $h_1$, each decaying promptly to $b\bar{b}$ pairs~\cite{ATLAS:2016tzx,ATLAS:2018pvw,ATLAS:2025rfm}. This process leads to striking final state signatures containing four $b$-tagged jets accompanied by either one charged lepton and missing transverse energy {\sl(from $W$ production)} or two charged leptons {\sl (from $Z$ production)}. We perform a comprehensive scan of the TRSM parameter space, implementing all relevant theoretical constraints (perturbativity, unitarity, vacuum stability) and experimental bounds from collider searches and electroweak precision measurements. Finally, we assess the discovery potential of these exotic signatures at the LHC Run 3 with 13.6~\TeV center-of-mass energy and an integrated luminosity of 300~\fb$^{-1}$, as well as at the High-Luminosity LHC {\sl(HL-LHC)} with 3000~\fb$^{-1}$. 

The remainder of this paper is organized as follows. Section~\ref{sec:model} reviews the TRSM framework and summarizes the theoretical and experimental constraints applied to the parameter space. Section~\ref{sec:Analyses} describes our simulation and analysis strategy, including event generation, detector simulation, and signal-to-background discrimination. Section~\ref{sec:results} presents our results for benchmark scenarios at both Run 3 and HL-LHC luminosities. Finally, Section~\ref{sec:conclusions} summarizes our findings and discusses future prospects. Some preliminary results of this work have previously been presented in the conference proceedings of Ref.~\cite{Desai:2026fvt}.

\section{TRSM Framework}\label{sec:model}
We concentrate here on the extension of the SM scalar sector by the addition of two real gauge-singlet scalar fields. Such models have been widely discussed in the literature~
\cite{Barger:2008jx,Alexander-Nunneley:2010tyr,Coimbra:2013qq,Ahriche:2013vqa,Costa:2014qga,Costa:2015llh,Ferreira:2016tcu,Chang:2016lfq,Muhlleitner:2017dkd,Dawson:2017jja,Chiang:2017nmu,Robens:2019kga,Robens:2022nnw,aali:2020tgr,Lane:2024vur, Grzadkowski:2018nbc, Curtin:2014jma, Kotwal:2016tex, Beniwal:2017eik, Cheng:2018ajh, Ghorbani:2019itr}
In our scenario, we impose an additional $\mathbb{Z}_2^S\,\otimes\,\mathbb{Z}_2^X$ symmetry that reduces the number of free parameters in the potential. Furthermore, we assign a vacuum expectation value to each field such that the additional symmetry is softly broken. This model commonly referred to as the \emph{Two Real Singlet Model}  (TRSM) \cite{Robens:2019kga,Robens:2022nnw,Robens:2025tew},
constitutes one of the minimal renormalisable extensions of the SM Higgs sector, while allowing for a rich scalar spectrum and non-trivial Higgs phenomenology.

A discrete $\mathbb{Z}_2^S \otimes \mathbb{Z}_2^X$ symmetry,
\begin{equation}
S \rightarrow -S \,, \qquad X \rightarrow -X \,,
\end{equation}
is imposed to forbid linear and cubic terms in the singlet fields. All other fields transform evenly under the symmetry. The most general renormalizable scalar potential involving the SM Higgs doublet $\Phi$ and two real singlet fields $S$ and $X$ and invariant under the $\mathbb{Z}_2^{S} \otimes
\mathbb{Z}_2^{X}$ symmetry group is then given by

\begin{align}
V(\Phi,S,X) =\;&
-\mu_\Phi^2\, \Phi^\dagger \Phi
+ \lambda_\Phi (\Phi^\dagger \Phi)^2
+ \frac{1}{2}\mu_{S}^2 S^2
+ \frac{1}{2}\mu_X^2 X^2 \nonumber \\
&+ \frac{1}{4}\lambda_{S} S^4
+ \frac{1}{4}\lambda_X X^4
+ \frac{1}{2}\lambda_{S X} S^2 X^2 \nonumber \\
&+ \frac{1}{2}\lambda_{\Phi S} (\Phi^\dagger \Phi) S^2
+ \frac{1}{2}\lambda_{\Phi X} (\Phi^\dagger \Phi) X^2 \, .
\end{align}
It has nine real parameters,
\[
\mu_\Phi^2,\ \lambda_\Phi,\ \mu_S^2,\ \mu_X^2,\ \lambda_S,\ \lambda_X,\ \lambda_{S X},\ 
\lambda_{\Phi S},\ \lambda_{\Phi X}.
\]

As we consider the case where all scalar fields acquire non-zero vacuum expectation values (VEVs), and expanding the fields around their VEVs, the CP-even gauge eigenstates $\phi_h$, $\phi_S$, and $\phi_X$ are obtained as
\begin{equation}
\Phi =
\begin{pmatrix}
0 \\
\dfrac{v + \phi_h}{\sqrt{2}}
\end{pmatrix}, \qquad
S = \frac{v_S + \phi_S}{\sqrt{2}}, \qquad
X = \frac{v_X + \phi_X}{\sqrt{2}} \, .
\end{equation}
As the singlets do not couple to electroweak gauge bosons in the gauge eigenstates, they do not contribute to electroweak symmetry breaking. Therefore, the VEV of the doublet is fixed by electroweak precision measurements to be  $v_\Phi = v_{\text{SM}}\,\sim\,246$ GeV. The VEVs of the singlet fields break the above symmetry,  resulting in a mixing of scalar field into the physical states according to

\begin{equation}
(h_1, h_2, h_3)^T = \mathcal{R} \, (\phi_h, \phi_S, \phi_X)^T,
\end{equation}
where $\mathcal{R}$ is a  standard unitary rotation matrix parametrized according to \cite{Robens:2019kga},
\begin{equation}
    \mathcal{R} = \begin{pmatrix}
        c_1 c_2             & -s_1 c_2             & -s_2     \\
        s_1 c_3-c_1 s_2 s_3 & c_1 c_3+ s_1 s_2 s_3 & -c_2 s_3 \\
        c_1 s_2 c_3+s_1 s_3 & c_1 s_3-s_1 s_2 c_3  & c_2 c_3
    \end{pmatrix}
\end{equation}

We introduce three mixing angles, $\theta_{hS}$, $\theta_{hX}$, and $\theta_{SX}$, and use the shorthand notation
$s_1\equiv\sin\theta_{hX}$,
$s_2\equiv\sin\theta_{hS}$,
$s_3\equiv\sin\theta_{SX}$,
$c_1\equiv\cos\theta_{hX}$,
$c_2\equiv\cos\theta_{hS}$, and
$c_3\equiv\cos\theta_{SX}$.
In order to comply with current measurements, one of the three scalar mass eigenstates $h_i$ ($i = 1,2,3$) needs to be identified with the observed SM-like Higgs boson with mass $m_h \simeq 125~\text{GeV}$. Note we also apply a mass hierarchy such that
\begin{\eqn*}
M_1\,\leq\,M_2\,\leq\,M_3,
\end{\eqn*}
where $M_i$ denotes the mass of the scalar $h_i$.

In the scenario considered here, we identify $h_3\,\equiv\,h_{125}$. As a result, the model is described by the remaining seven independent real parameters, which we choose as:
\begin{equation}
    M_1, M_2, \theta_{hS}, \theta_{hX}, \theta_{SX}, v_S, v_X,
\end{equation}
The parameter space of the TRSM discussed above is subject to several theoretical and experimental constraints. Theoretical constraints include perturbative unitarity and the requirement that the scalar
potential be bounded from below~\cite{Robens:2019kga, Kannike:2012pe, Kannike:2016fmd, Ferreira:2016tcu}. Experimental constraints arise from Higgs signal strength measurements, direct searches for
additional scalar resonances, electroweak precision observables ($S$, $T$, and $U$~\cite{Altarelli:1990zd, Peskin:1990zt, Peskin:1991sw}), as well as limits on non-standard and invisible Higgs decays. These constraints are implemented using the \texttt{ScannerS} framework~\cite{Coimbra:2013qq,Muhlleitner:2020wwk} and \texttt{HiggsTools}~\cite{Bahl:2022igd,Bahl:2026yal}\footnote{Results were checked in August 2026.}. 
We refer the reader to \cite{Robens:2019kga,Robens:2022nnw,Robens:2025tew} for a more detailed discussion of the parameter scan\footnote{See also \cite{Robens:2026dwz} for a tool that maximizes rates for specific final states within this model.}.

\subsection{Phenomenology}
The extended scalar sector of the TRSM gives rise to a rich and diverse phenomenology.  In particular, the presence of multiple scalar states enables cascade decays,  modifies Higgs self-couplings, and allows for exotic decay modes of the multi-scalars Higgs boson. To guide and facilitate dedicated experimental searches, six benchmark scenarios have been proposed within the TRSM framework~\cite{Robens:2019kga,Robens:2022nnw}. Each benchmark is designed to highlight one or more distinctive signatures and is chosen to maximise the expected signal yield, while remaining consistent with theoretical and experimental constraints.

We focus on \texttt{BP4}, in which $h_3$ is identified as the SM-like Higgs boson and $h_2 \to h_1 h_1$. We perform a random scan over the TRSM parameter space within the ranges specified in Table~\ref{tab:parametrs}.
\begin{table}[H]
    \centering
    \begin{tabular}{c|ccccccccc}
    \hline\hline
         Parameters&$M_{1}$&$M_{2}$&$M_{3}$&$\theta_{hs}$&$\theta_{hX}$&$\theta_{SX}$&$v_{\Phi}$&$v_{S}$&$v_{X}$  \\\hline
         Ranges&$[1,\ 62]$&$[1,\ 124]$&$125.09$&$-1.284$&$1.309$&$-1.519$&$v_{\text{SM}}$&$990$&$310$ \\\hline\hline
    \end{tabular}
    \caption{Parameters in TRSM. Masses and VeVs are in GeV. 
    }
    \label{tab:parametrs}
\end{table}

While the above table lists the input parameters or free parameters, one important quantity are the rescaling factors

\begin{\eqn*}
\kappa_i\,\equiv\,R_{i1}
\end{\eqn*}

with $R$ denoting the mixing matrix introduced above. These rescaling factors determine the coupling of the scalar $h_i$ for decays that are purely inherited from the SM like doublet such that

\begin{\eqn*}
g_{h_i x\,y}\,=\,\kappa_i\,g_{h x\,y}^\text{SM}
\end{\eqn*}

where $g$ is the coupling strength for the respective decay into particles $x,\,y$ and $g^\text{SM}$ the corresponding coupling in the SM. These factors are constant across the BP4 plane and have been calculated as \cite{Robens:2019kga}

\begin{figure}[htb]
\centering
\includegraphics[scale=0.45]{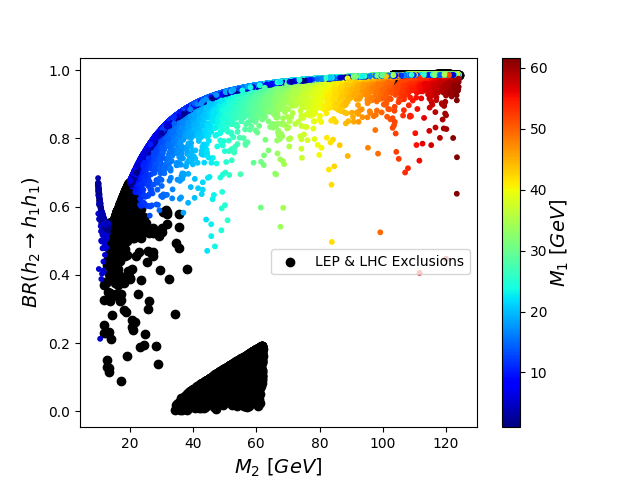}
\includegraphics[scale=0.45]{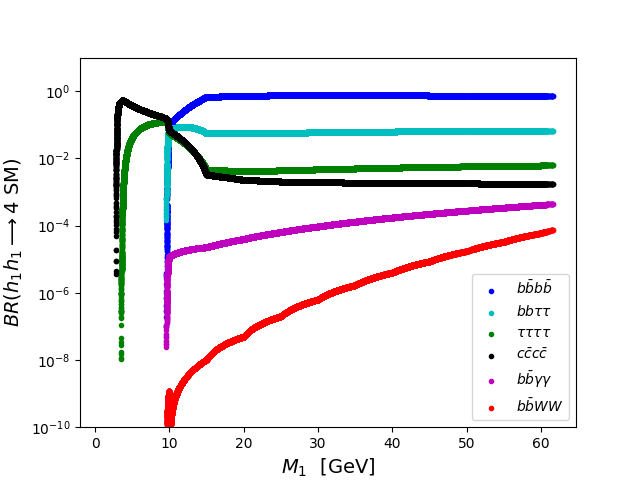}
\caption{Branching ratios $\mathrm{BR}(h_2 \to h_1 h_1)$ as functions of $M_1$ and $M_2$, and $\mathrm{BR}(h_1 h_1 \to 4\ \textrm{SM})$ as a function of $M_1$ {\sl (in~\GeV)}.} \label{figs:BR_h1h1_1}
\end{figure}

\begin{\eqn*}
\kappa_1\,=\,0.073,\,\kappa_2 \,=\,0.223, \, \kappa_3\,=\,0.972
\end{\eqn*}

respectively.
In Figure~\ref{figs:BR_h1h1_1}, we show the branching ratios of the decay chains $h_2 \to h_1 h_1$ {\sl (left)} and $h_1 h_1 \to 4$SM {\sl (right)} for the parameter space
defined in Table~\ref{tab:parametrs}. The branching ratio $\mathrm{BR}(h_2 \to h_1 h_1)$ varies between 0.8 and 1.0 for $M_1$ in the range 20--60~GeV. For the decay $h_1 h_1 \to 4$SM, the $b\bar b$ channel dominates in the region where $M_1\,\gtrsim\,20\,\GeV$, followed by the $\tau^+\tau^-$ and $c\bar c$ final states. This behaviour is consistent with our analysis strategy, which focuses on four-$b$ final states.

\subsection{$Vh_2$ Production}
The associated production cross sections \(pp \to W^\pm h_i\) and \(pp \to Z h_i\)  are obtained by rescaling the corresponding SM cross sections  according to the modified Higgs--gauge boson couplings. The coupling of a scalar mass eigenstate \(h_i\) to electroweak gauge bosons  is proportional to its doublet component, quantified by the mixing matrix element \(R_{i1}\). Consequently, the production cross section is given by
\begin{equation}
\sigma(pp \to V h_i) = \kappa_i^2 \, 
\sigma_{\rm SM}(pp \to Vh)\big|_{m_h = m_{h_i}},
\end{equation}
with \(\kappa_i = R_{i1}\). The SM reference cross sections are taken from the LHC Higgs Cross Section Working Group and include higher-order QCD and electroweak corrections, evaluated at NNLO accuracy for a proton--proton center-of-mass energy of \(\sqrt{s} = 13~\mathrm{TeV}\)~\cite{LHCHiggsCrossSectionWorkingGroup:2016ypw}. This approach assumes on-shell Higgs production within the narrow-width approximation in a factorized approach and neglects off-shell effects and interference contributions.
Note that we will present a phenomenological study for LHC Run 3, i.e. with a center of mass energy of 13.6 \TeV. We however assume the scaling between 13 \TeV and 13.6 \TeV to be roughly independent of the mass of the produced particle, therefore, 13 \TeV cross sections should suffice to determine interesting benchmark scenarios.

The \(W^\pm h_2\) and \(Z h_2\) associated production processes are treated separately, as their production cross sections differ and their experimental analyses at colliders often follow distinct strategies. These cross sections are calculated within the \texttt{ScannerS} framework~\cite{Coimbra:2013qq,Muhlleitner:2020wwk}.

\begin{figure}[htb]
 	\centering
 	\includegraphics[scale=0.55]{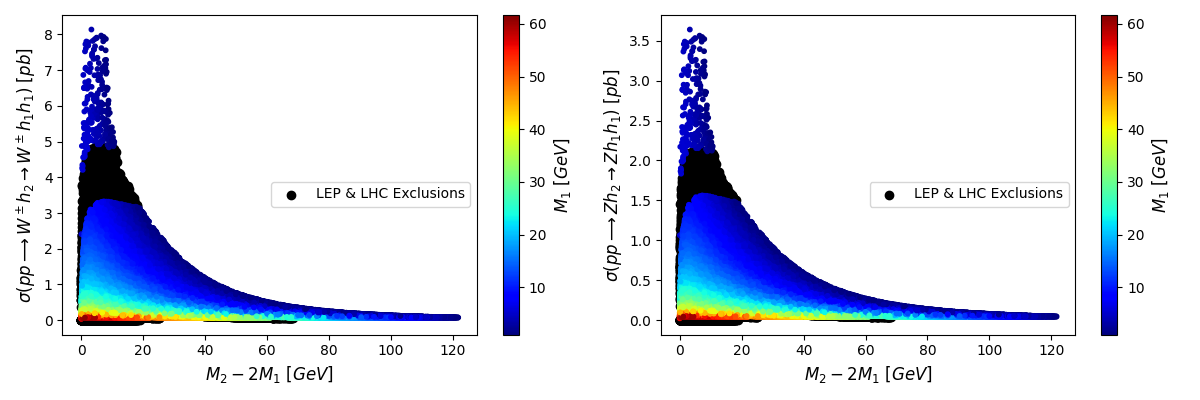}
    \caption{Predicted production cross section times branching ratio, $\sigma \times \mathrm{BR}(h_2 \to h_1 h_1)$, at $\sqrt{s}=13$~\TeV for the $W^\pm h_2$ {\sl (left)} and $Zh_2$ {\sl (right)} production channels, shown as a function of $(M_2-2M_1,\;M_1)$. Black points indicate parameter points excluded by the experimental constraints.}
\label{figs:xs_and_br}
 \end{figure}

Figure~\ref{figs:xs_and_br} presents the production cross sections \(\sigma(pp \to W^\pm h_{2} \to W^\pm h_{1} h_{1})\) {\sl (in~\pb, left)} and \(\sigma(pp \to Z h_{2} \to Z h_{1} h_{1})\) {\sl (in~\pb, right)} as functions of the mass parameters  {\sl \((M_{1},\, M_{2} - 2M_{1})\)}. These production cross sections remain sizeable in the kinematic region $M_{2} \approx 2M_{1}$, particularly for $M_{2}-2M_{1} \leq 40~\mathrm{GeV}$. 

To evaluate the discovery prospects of light scalar states at the LHC and the HL-LHC, we consider a representative set of benchmark points (BPs), listed in Table~\ref{tab:Bps-trsm}, together with their corresponding model parameters, production cross sections, branching ratios, and decay widths. These benchmarks are selected to probe different regions of the TRSM parameter space, including compressed spectra with $M_2 \simeq 2M_1$, intermediate-mass scenarios that provide optimal signal sensitivity, and high-$M_2$ regimes with $M_2\,\sim\,120\,\GeV$ where the production cross sections become significantly suppressed. These benchmark values are obtained using \texttt{ScannerS}~\cite{Muhlleitner:2020wwk}. The benchmark points are illustrated in Fig.~\ref{bps_m1m2xs}. In the left panel, the BPs are shown in the plane of the scalar masses $(M_1,\ M_2)$, with the color scale representing the production cross section $\sigma(pp \to Vh_2 \to Vh_1 h_1)$ in \pb. The right panel displays the same benchmark points in the $(M_2 - 2M_1,\ \sigma(Vh_1h_1) )$ plane, with color-coded according to $M_1$. The cross sections, decay widths, and branching ratios used in the event simulation are calculated with \texttt{MadGraph5\_aMC@NLO}~\cite{Alwall:2014hca} using the corresponding TRSM model implementation\footnote{\url{https://gitlab.com/apapaefs/twosinglet}}\cite{Papaefstathiou:2020lyp}. The resulting values are summarized in Table~\ref{tab:combined} and, for completeness, in Table~\ref{tab:MG5-withsBR} of Appendix~\ref{withs-MG5}. Consequently, small differences may arise between the benchmark values obtained with \texttt{ScannerS} and those used in the event simulation. The values tabulated in \texttt{ScannerS} include higher order contributions via \texttt{HDecay}~\cite{Djouadi:2018xqq}, while the values from \texttt{MadGraph\_aMC@NLO}~\cite{Alwall:2014hca} are performed at tree level. For internal consistency we decided to use the latter in our calculations.
{\renewcommand{\arraystretch}{1.4} 
{\setlength{\tabcolsep}{0.1cm} 
\begin{table}[htb!]
\centering
\begin{tabular}{c|ccccccccc}
\toprule
  &$M_{1}$ & $M_{2}$ & $M_{2}-2M_{1}$ & $\sigma(Z h_2)$ & $\sigma(W^\pm h_2)$ & $Br(h_2\to h_1 h_1)$ &$Br(h_1\to b \bar{b})$& $\Gamma_{h_1}(10^{-6})$ & $\Gamma_{h_2}(10^{-3})$ \\
\midrule
 BP1&20.13 & 42.96  & 2.7  & 0.63 & 1.27 & 0.87&0.856& 3.31 & 0.46 \\
 BP2&24.84 & 55.25  & 5.56 & 0.37 & 0.71 & 0.93&0.866& 4.04 & 1.1 \\
 BP3&21.6 & 58.31 & 15.12  & 0.33 & 0.63 & 0.95&0.859& 3.53 & 1.6 \\
 BP4&30.16 & 70.16  & 9.83  & 0.21 & 0.39 & 0.96& 0.869& 4.77 & 2.42 \\
 BP5&22.27 & 79.57  & 35.03 & 0.16 & 0.28 & 0.98&0.860 & 3.64 & 4.03 \\
 BP6&36.7 & 79.95  & 6.54 &  0.15 & 0.28 & 0.97&0.867& 5.63 & 2.98 \\
 BP7&37.83 & 90.12  & 14.46 & 0.11 & 0.2 & 0.98& 0.867& 5.78 & 5.23 \\
 BP8&20.92 & 98.23  & 56.38 & 0.09 & 0.15 & 0.98& 0.857& 3.43 & 7.31 \\
 BP9&20.58 & 118.72 & 77.57  & 0.05 & 0.09 & 0.99& 0.857& 3.38 & 12.59 \\
\bottomrule
\end{tabular}
\caption{BPs selected for the TRSM study. The corresponding masses, decay widths, production cross sections at $13$ \TeV, and branching ratios are obtained using \texttt{ScannerS} for the parameter sets listed in Table~\ref{tab:parametrs}. Masses and widths are given in \GeV, and cross sections in \pb. 
}
\label{tab:Bps-trsm}
\end{table}

\begin{figure}[htb!]
    \centering
    \includegraphics[width=0.95\linewidth]{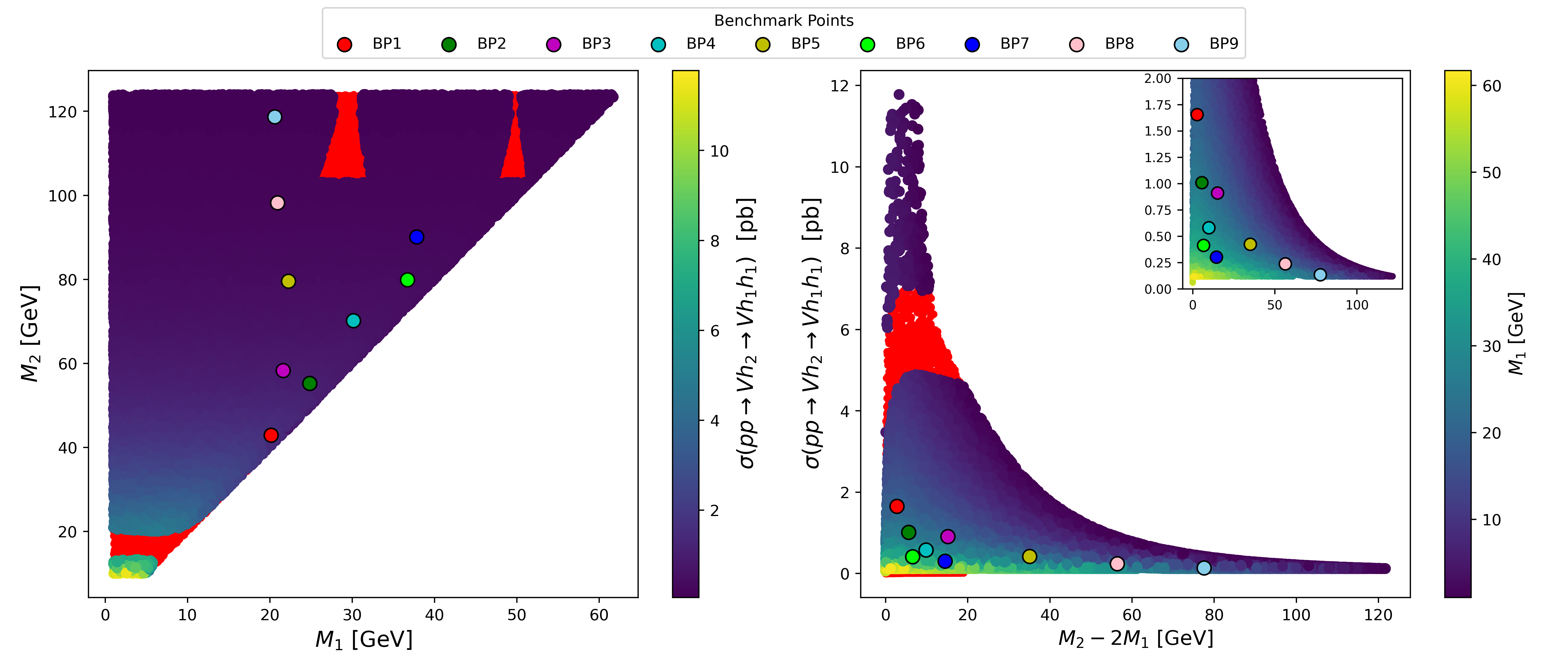}
\caption{BPs are shown in the $(M_{1},\, M_{2})$ plane, together with the production cross sections $\sigma(p p \to Vh_2\to Vh_1h_1)$[\pb], in the left panel.
The right panel displays the same points in the $(M_{2} - 2M_{1},\ M_1,\, \sigma)$ plane.  The red-shaded regions indicate the portions of parameter space excluded by current collider constraints, see also \cite{Robens:2025tew}.}\label{bps_m1m2xs}
\end{figure}

\section{Signal and Background Analysis}\label{sec:Analyses}
\subsection{Signal}
This study focuses on the $Wh_2$ and $Zh_2$ productions processes, with $W \rightarrow \ell \nu_\ell$, $Z\rightarrow \ell^+ \,\ell^-\ (\ell=e,\ \mu)$ and $h_2 \rightarrow h_1 h_1 \rightarrow 4b$, as shown by the Feynman diagram depicted in Figure \ref{fig:Feynman_diagram} at tree level. The $Zh_2$ associated production receives also a non-negligible contribution from loop induced $g g \to Zh_2$. The resultant signature contains either one or two leptons, electron or muon together with four b-quarks (b-jets). The presence of charged leptons in the final state offers a robust signature for triggering and effectively suppressing background contributions arising from the high cross-section strong-interaction production of four b-jets.
In this work, we investigate the BPs in table~\ref{tab:Bps-trsm}, taking signal and background into account in a dedicated analysis for each point. Events for both signal and background processes are generated using \texttt{MadGraph5\_aMC@NLO}~\cite{Alwall:2014hca}. The decays of unstable particles in signal and background samples are handled by \texttt{MadSpin}~\cite{Artoisenet:2012st}, ensuring the preservation of spin correlations between the final-state particles and their corresponding parent resonances. Signal and background cross sections are evaluated using the \texttt{NNPDF23\_nlo\_as\_0119} parton distribution functions (PDFs).
The renormalization and factorization scales are set dynamically on an event-by-event basis and are defined as the sum of the transverse momenta of all final-state particles, $\mu_R = \mu_F = \sum_{i} m_{T,i} / 2$. 

\begin{figure}[htb!]
    \centering
    \includegraphics[width=0.4\linewidth]{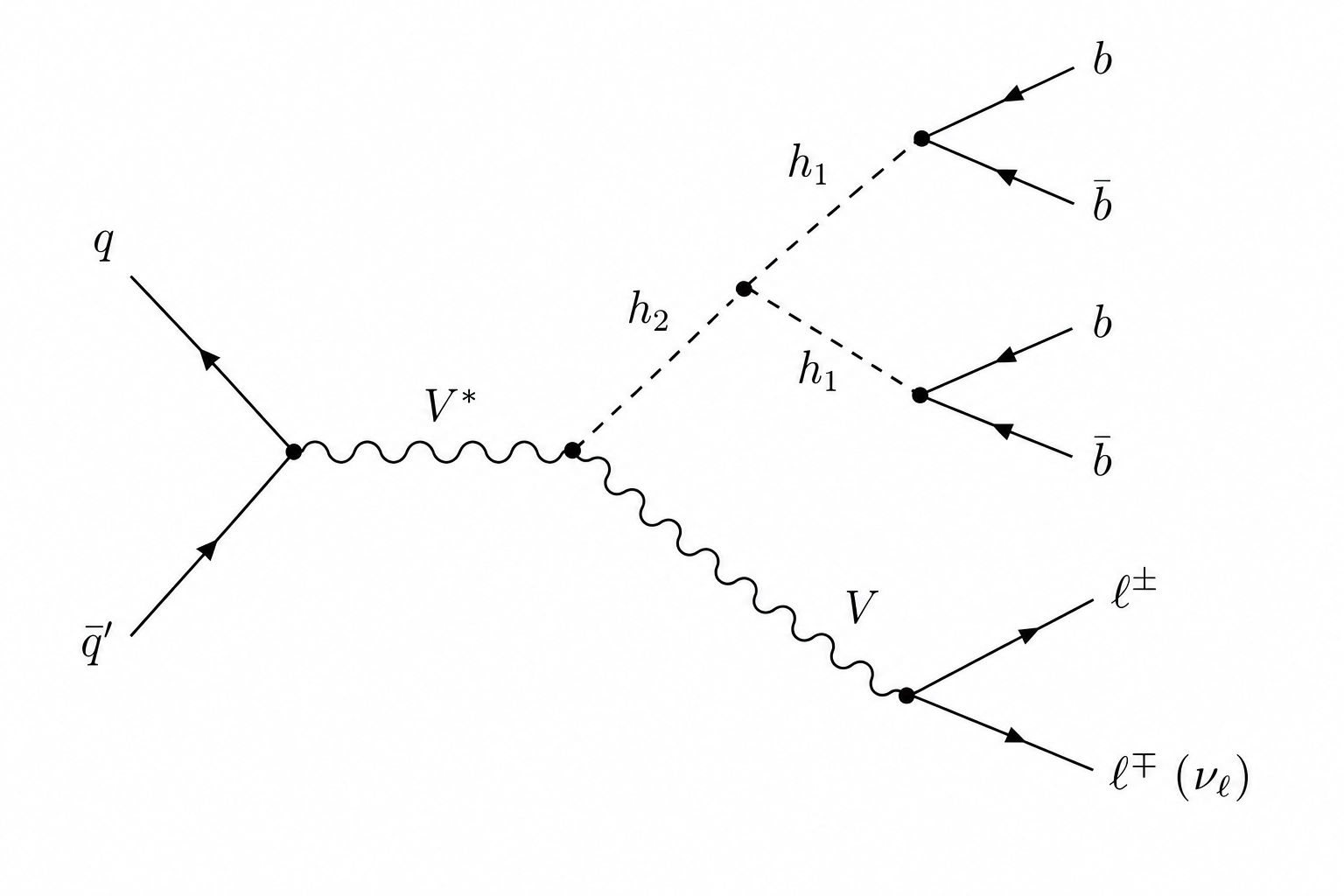}
    \caption{Tree-level Feynman Diagram for the target process, were $V\,\equiv\,W\,\text{or}\, Z$}
    \label{fig:Feynman_diagram}
\end{figure}

Parton showering and hadronisation are performed with \texttt{Pythia~8}~\cite{Sjostrand:2014zea}, where for the leading-order event samples, the matching between the matrix-element calculations and the parton shower is implemented using the MLM prescription~\cite{Alwall:2014hca}. The detector response is simulated using \texttt{Delphes}~\cite{deFavereau:2013fsa} with a CMS detector card. Jet reconstruction for both signal and background events is performed using \texttt{FastJet}~\cite{Cacciari:2011ma} with the anti-$k_T$ algorithm~\cite{Cacciari:2008gp} and a cone radius of $ R = 0.4$. Transverse momentum ($p_T$) requirements are imposed on light-flavor jets, heavy-flavor jets, and leptons. Events are retained only if these objects satisfy the following selection criteria:

\begin{align}
    p_T(j) \ge 20 \GeV,\ p_T(b) \ge 20 \GeV,\ p_T(\ell) \ge 10 \GeV\label{based_cuts1}
\end{align}

Additional selection cuts are imposed on the pseudorapidity and angular separation between particles in the reconstructed events, following
\begin{align}
    |\eta | < 2.5,\ \Delta R (i,\ j) \geq 0.4, \ i,j = b, j, \ell\label{based_cuts2}
\end{align}

{
The identification of $b$-jets is modeled according to the \texttt{DeepCSV} algorithm at the medium working point, using the parametrized efficiencies provided in Ref.~\cite{CMS:2017wtu}, scaled to reflect the improvements reported in Ref.~\cite{CMS-DP-2024-066}. These parametrizations supply the $b$-tagging efficiency as a function of the jets transverse momentum for the different operating points of the \texttt{DeepCSV} algorithm, and are well suited for phenomenological studies relying on $b$-jet identification. In this analysis, we adopt the medium working point parametrization without implementing the full machine-learning algorithm. This corresponds to an average $b$-tagging efficiency of about $70-75\%$, a charm-jet misidentification rate of approximately  $\varepsilon_{c\to b} \simeq 1-2\%$, and a light-flavor ($u,d,s,g$) jet misidentification rate of about $\varepsilon_{j\to b} \simeq 0.1-0.2\%$. These efficiencies are implemented in Delphes through $p_T$-dependent parameterizations, ensuring a realistic modelling of flavour-tagging performance across the relevant kinematic range $[20,\ 1000]$ \GeV. The analysis of the events at the end was done with \texttt{MadAnalysis5} framework~\cite{ma5:2012fm}, and a python based script.}

\subsection{Background Processes}
\label{sec:bgd}
As discussed above, our main channel for the single lepton case is $Wh_2$ production and is characterized by one lepton, missing transverse momentum, and multiple $b$-tagged jets. Therefore, the main final state corresponds to a lepton, $4 b$, as well as missing transverse energy. As we will discuss below, in our analysis we however require at least 2 tagged b-jets. This gives rise to a large number of backgrounds, including those where light jets are mistagged as b-jets and vice versa. Following~\cite{ATLAS:2018pvw,ATLAS:2025rfm}, we identify the dominant background as arising from semileptonic $t\bar t+jets$ production, and $W$+jets processes with heavy-flavor content. Additional smaller contributions arise from single-top ($tW$), diboson production ($WW$, $WZ$) and associated top--vector-boson processes ($t\bar{t}V$). The dilepton channel targets $Zh_2$ production and requires two same-flavor, opposite-sign leptons consistent with a $Z$ boson decay, together with multiple $b$-tagged jets; in this case, the dominant background is $Z$+jets production, in particular $Z+b\bar b$, with additional subdominant contributions from dileptonic $t\bar t+$jets and diboson processes.  Table~\ref{tab:backgrounds} lists the dominant and subdominant background processes considered in this analysis. Their corresponding leading-order (LO) cross sections at $\sqrt{s}=13.6$~\TeV, after detector simulation and the baseline selection defined in Eqs.~\eqref{based_cuts1}--\eqref{based_cuts2}, are reported in Table~\ref{tab:xs_backgrounds}.\footnote{Appendix~\ref{ir_BG} summarizes the irreducible background processes and their corresponding contributions after the full event selection. In the main analysis, only the dominant and subdominant background processes listed in Table~\ref{tab:backgrounds} are retained for each benchmark point, since all remaining backgrounds are found to have a negligible impact on the final sensitivity. In addition, the corresponding $K$-factors for the signal and background processes are provided in Tables~\ref{tab:without_decay} and~\ref{tab:kfactors}.} 
\begin{table}[H]
\centering
\setlength{\tabcolsep}{6pt}
\renewcommand{\arraystretch}{1.3}
\begin{tabular}{l|cccccc}
\toprule
Channel & $t\bar{t}$+jets & Single top & $W$+jets & $Z$+jets & Diboson & $t\bar{t}V$ \\
\midrule
Single-Lepton ($W h_2$) & \checkmark & \checkmark & \checkmark & -- & \checkmark & \checkmark \\
Dilepton ($Z h_2$) & \checkmark & \checkmark & -- & \checkmark & \checkmark & \checkmark \\
\bottomrule
\end{tabular}
\caption{Main background processes for the single-lepton and dilepton channels.}
\label{tab:backgrounds}
\end{table}

\begin{table}[H]
    \centering
    \setlength{\tabcolsep}{6pt}
    \renewcommand{\arraystretch}{1.1}
    \begin{tabular}{llccc}
        \toprule
        & & \multicolumn{3}{c}{Cross-section [\pb]} \\
        \cmidrule(lr){3-5}
        Process & Decay channel & $W^+$ & $W^-$ & $Z$ \\
        \midrule
         \multirow{3}{*}{$t\bar{t}$+0,1,2,jets} & $t\bar{t}\to b\bar{b}W^+W^- \to b\bar{b}\ell^+ \nu_\ell jj$ & $89.78 (4)$ & -- & -- \\
            & $t\bar{t}\to b\bar{b}W^+W^-\to b\bar{b}jj\ell^- \bar{\nu}_\ell$ & -- & $89.78 (4)$ & -- \\
     & $t\bar{t}\to b\bar{b}W^+W^- \to b\bar{b}\ell^+ \nu_\ell \ell^-\bar{\nu_\ell}$ & -- & -- & $29.89 (1)$ \\
        \midrule
     $W$+1,2\,jets & $W \to \ell \nu_\ell$ & $475.3 (6)$ & $323.8 (5)$ & -- \\
        \midrule
     $Z$+1,2\, jets & $Z \to \ell^+ \ell^-$ & -- & -- & $812.4 (4)$ \\
     $tW$ &$t\to Wb$& $0.02415 (1)$&$0.02846(1)$&--\\
     $WW$&$W\to \ell\nu_\ell(jj)$&$11.39(1)$&$11.39(1)$&$3.795(2)$\\
     $WZ$&$W(Z)\to \ell\nu_\ell(jj\ \textrm{or}\ \ell^+\ell^-)$&$2.757(2)$&$1.751(1)$& $1.342(1)$\\
     $ZZ$&$ZZ\to jj\ell^+\ell^-$&--&--&$3.308(2)$\\
     $ttV$&$t\to bW$&$0.05525(5)$&$0.0285(3)$&$0.04201(4)$\\
     $W4b$&$W\to \ell\nu_\ell$&$0.09588(6)$&$0.06057(5)$&$-$\\

        \bottomrule
    \end{tabular}
    \caption{Leading-order cross sections for background processes at $\sqrt{s}=13.6$ \TeV after detector simulation. 
    }
    \label{tab:xs_backgrounds}
\end{table}

\section{Results and discussion }
\label{sec:results}

Table \ref{tab:combined} lists the production cross sections for the target process
\begin{\eqn*}
p\,p\,\rightarrow\,h_2\,W^\pm(Z)\,\rightarrow\,h_1\,h_1\,\ell\,\nu_\ell (\ell^+\ell^-)
\end{\eqn*}

with subsequent decays into $b\,\bar{b}$ for the $h_1$, where based cuts defined in Eqs.~\eqref{based_cuts1}--\eqref{based_cuts2}.  were applied.  Decay widths calculated using Madgraph are also listed. 

\renewcommand{\arraystretch}{1.1}
\setlength{\tabcolsep}{0.06cm}
\begin{table}[H]
    \centering
    \begin{tabular}{|c|ccc|cccc|cc|}
        \hline
        \multirow{2}{*}{BP} & \multicolumn{3}{c|}{Masses [GeV]} & \multicolumn{4}{c|}{Cross-sections $\sigma$ [\pb]} & \multicolumn{2}{c|}{Widths(\MeV)} \\
        \cline{2-10}
        & $M_1$ & $M_2$ & $M_2-2M_1$ & $W^+h_2$ & $W^-h_2$ & $qq(Zh_2)$ & $gg(Zh_2)$ & $\Gamma_{h_1}\times 10^{-3}$ & $\Gamma_{h_2}$ \\
        \hline
BP1  & 20.13 & 42.96 & 2.70  & 0.10002 (7)  & 0.07028 (5)  & 0.02552 (2)  & 0.006252 (8) & 3.458  & 0.4943 \\
BP2  & 24.84 & 55.25 & 5.56  & 0.06299 (3)  & 0.04362 (3)  & 0.01625 (1)  & 0.005733 (6) & 4.861  & 1.145 \\
BP3  & 21.60 & 58.31 & 15.12 & 0.05661 (3)  & 0.03907 (3)  & 0.01465 (1)  & 0.005706 (5) & 3.906  & 1.650  \\
BP4  & 30.16 & 70.16 & 9.83  & 0.03672 (3)  & 0.02502 (2)  & 0.00962 (1)  & 0.005133 (4) & 6.361  & 2.484  \\
BP5  & 22.27 & 79.57 & 35.03 & 0.02648 (2)  & 0.01785 (1)  & 0.006983 (4) & 0.004795 (4) & 4.107  & 4.109  \\
BP6  & 36.70 & 79.95 & 6.54  & 0.02618 (2)  & 0.01765 (1)  & 0.006907 (4) & 0.004702 (5) & 8.131  & 3.060  \\
BP7  & 37.83 & 90.12 & 14.46 & 0.01918 (1)  & 0.01281 (1)  & 0.005095 (3) & 0.004371 (5) & 8.431  & 5.322  \\
BP8  & 20.92 & 98.23 & 56.38 & 0.01477 (1)  & 0.009768 (7) & 0.003937 (2) & 0.003745 (4) & 3.700  & 7.403  \\
BP9  & 20.38 & 118.72& 77.57 & 0.008427 (6) & 0.005465 (4) & 0.002262 (1) & 0.003356 (3) & 3.597  & 12.69  \\
        \hline
    \end{tabular}
    \caption{Leading-order cross-sections (in \pb) at the LHC with $\sqrt{s}=13.6$ \TeV, using Eqs.~(\ref{based_cuts1}) and (\ref{based_cuts2}). The signals correspond to $4b + \ell^+\nu_\ell$ ($W^+h_2$), $4b + \ell^-\bar{\nu}_\ell$ ($W^-h_2$), and $4b + \ell^+\ell^-$ ($qq$ and $gg\to Zh_2$). For all points,  $\Gamma_{h_3} = 6.201 $ \MeV.
    }
    \label{tab:combined}
\end{table}

In order to identify variables that might be helpful to select cuts, in Figure \ref{fig:wptab1}, we display the normalized distributions for various kinematic variables for the signal as well as backgrounds processes, where we have selected a few benchmark points.  We can see that several variables, as e.g. invariant mass or b-jet multiplicity in the $b\,\bar{b}$ system, could in principle work well to discriminate signal from background, at least for the nine benchmark points considered.

\begin{figure}[htb!]
	\centering
	\includegraphics[width=0.45\linewidth]{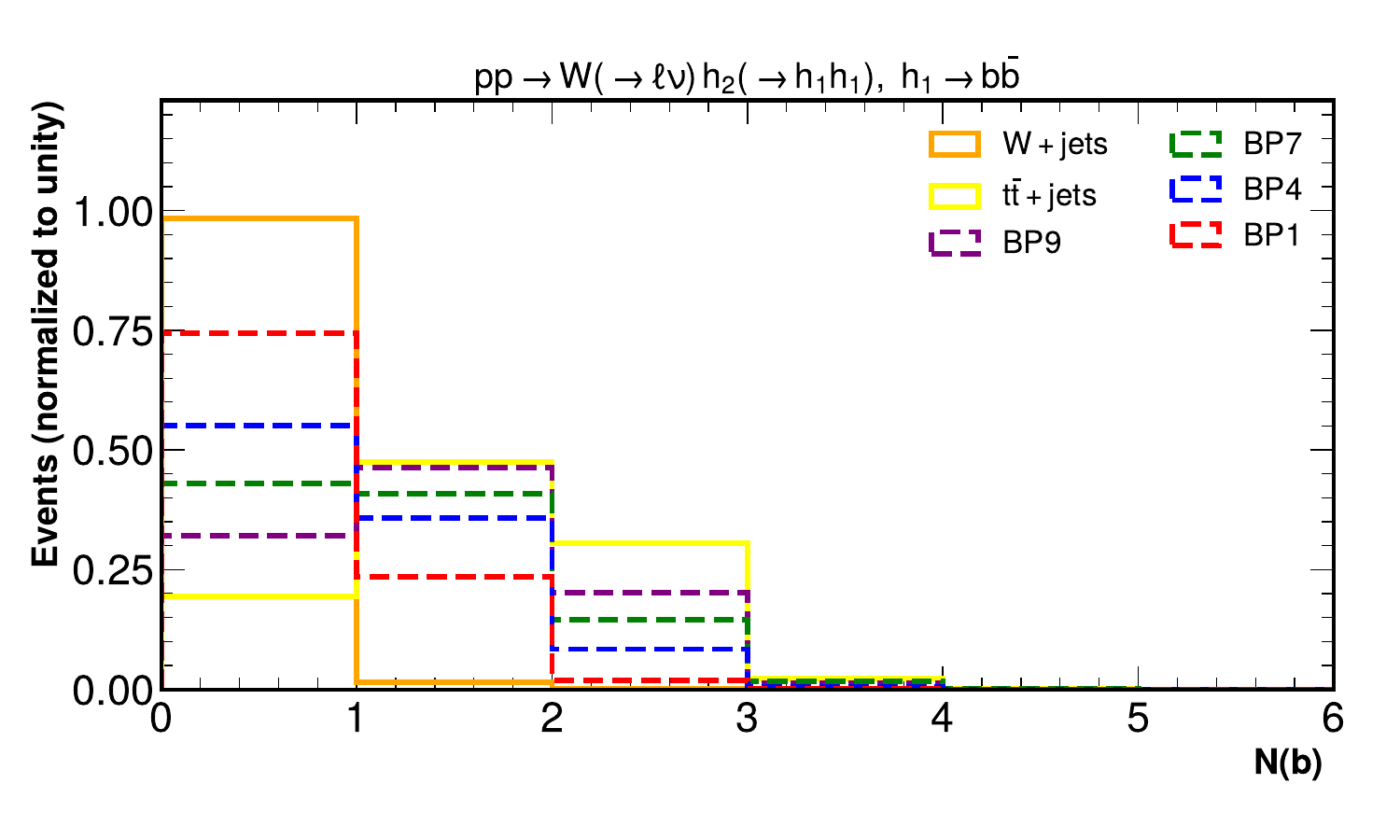}
	\includegraphics[width=0.45\linewidth]{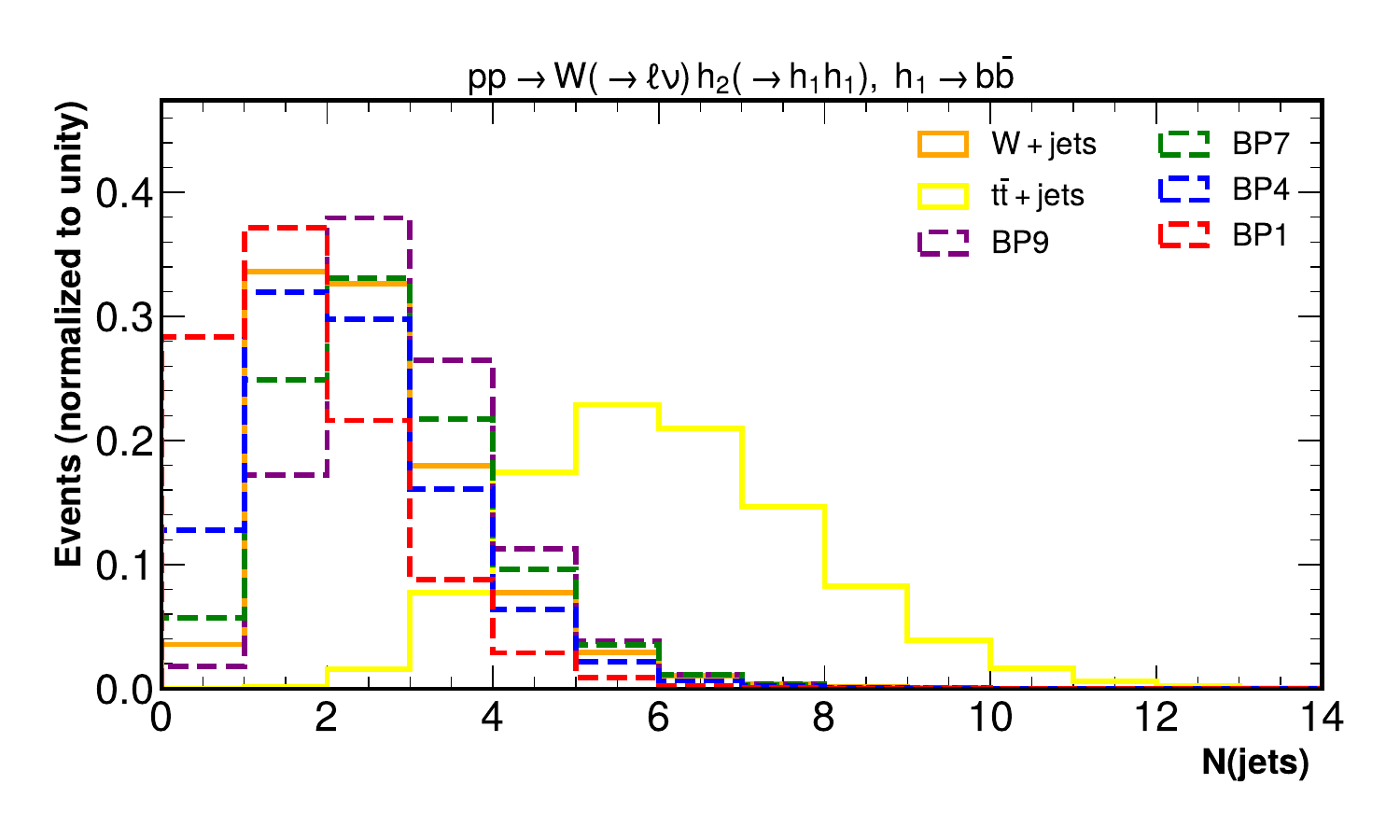}\\
	\includegraphics[width=0.45\linewidth]{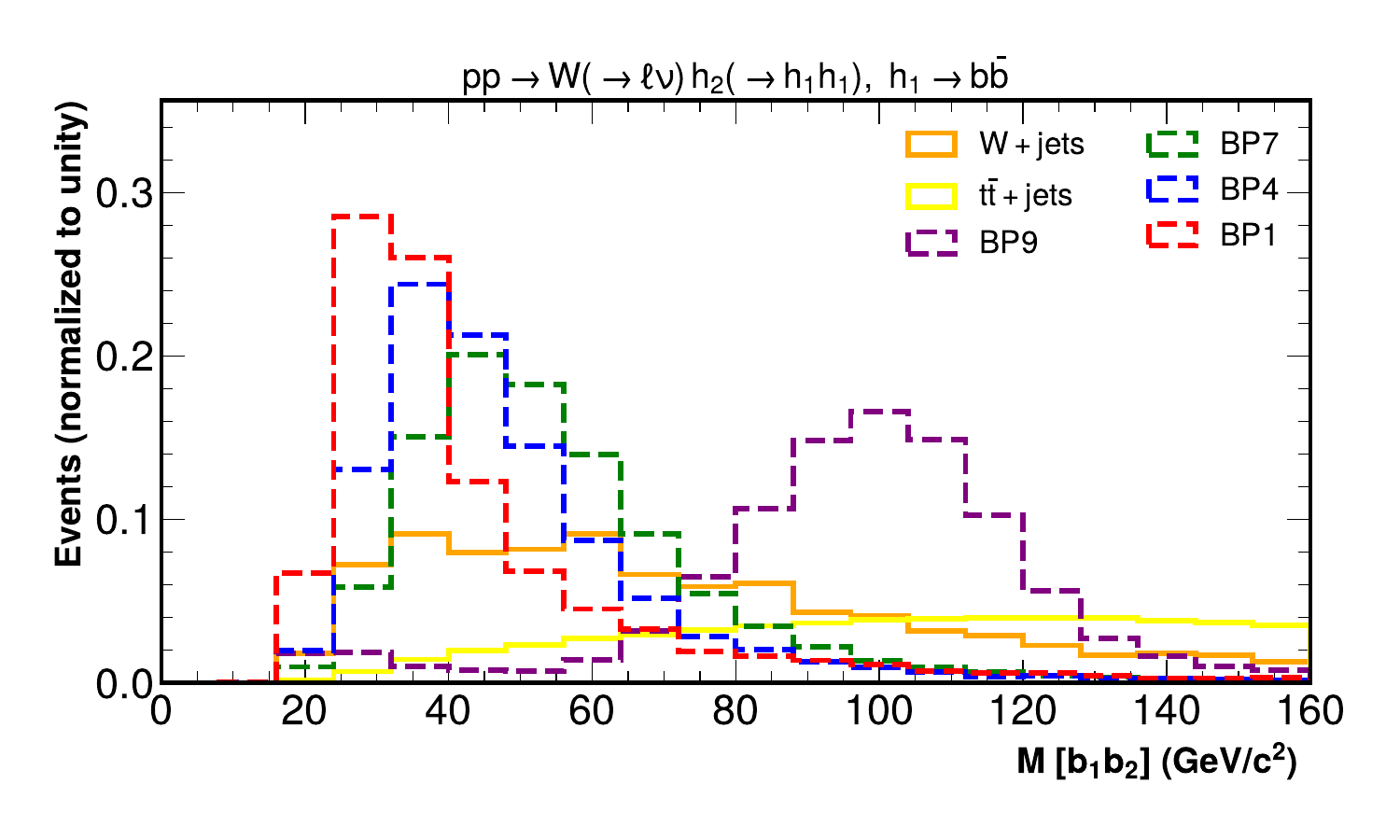}
	\includegraphics[width=0.45\linewidth]{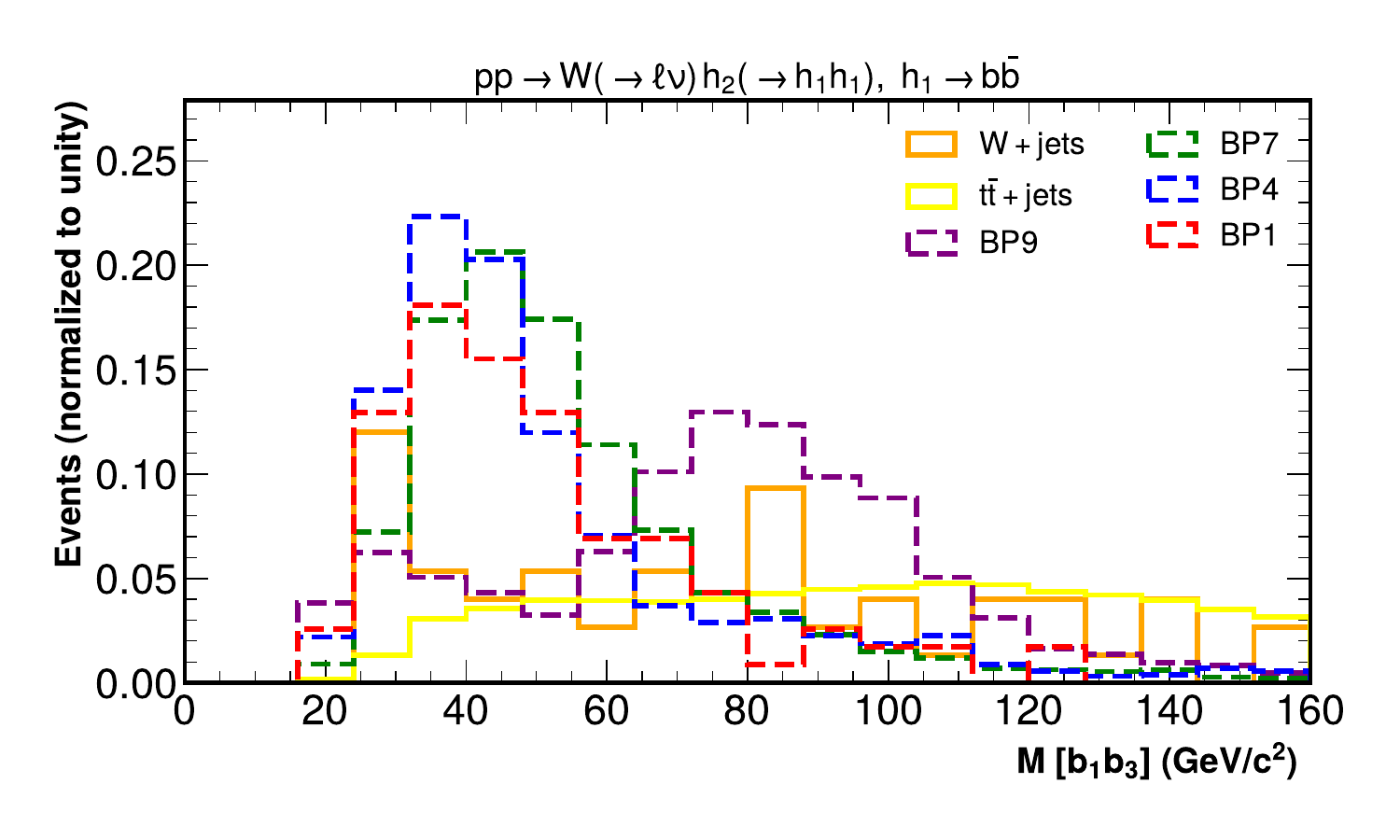}
	\caption{\label{fig:wptab1} Distributions of the number of $b$-jets ($N(b)$, {\sl upper left}), the number of light-jets ($N(jets)$, {\sl upper right}), the invariant mass of the leading and subleading $b$-jets ($M(b_1,b_2)$, {\sl lower left}), and the invariant mass of the leading and third-leading $b$-jets ($M(b_1,b_3)$, {\sl lower right}). The baseline selection cuts are defined in Eqs.~\eqref{based_cuts1}--\eqref{based_cuts2}. All distributions are normalized to unity for shape comparison.}
\end{figure}
Although the BP1 and BP9 have nearly identical light-scalar masses,
\[
M_{1}^{\rm BP1}=20.13~\mathrm{GeV},
\qquad
M_{1}^{\rm BP9}=20.38~\mathrm{GeV},
\]
their reconstructed $M(b_1,b_2)$ distributions after detector simulation can differ significantly as can be seen from the lower panel of Figure~\ref{fig:wptab1}. This behavior originates from the large difference in the parent scalar mass $M_{2}$, which strongly affects the kinematics of the $h_1$ decay products.

For BP1,
\[
M_{2}=42.96~\mathrm{GeV}\simeq 2M_{1},
\]
so the two $h_1$ scalar are produced nearly at threshold. Consequently, the $b$-quarks originating from each $h_1$ decay are relatively soft and less boosted, resulting in a topology where the correct $b\bar b$ pairing is more easily reconstructed, as we will discuss below.

In contrast, for BP9,
\[
M_{2}=118.72~\mathrm{GeV}\gg 2M_{1},
\]
the $h_1$ bosons are produced with a substantial boost. As a result, the two $b$-quarks from a given $h_1$ tend to become more collimated, with an approximate angular separation
\[
\Delta R(b,b)\sim \frac{2M_{1}}{p_T(h_1)}.
\]
This can lead to partial jet merging, reduced reconstruction efficiency, or the loss of one of the two $b$-jets due to detector and jet-selection requirements. In Figure~\ref{fig:deltaR_Mb1b2} we display distributions of $\Delta R(b_1,b_2)$ {\sl (upper panels)} and the invariant mass $M(b_1,b_2)$ {\sl (lower panels)} for BP1 and BP9. The left panels correspond to the parton level before jet clustering, whereas the middle and right panels show the hadron level after jet reconstruction with the anti-$k_T$ algorithm using radius parameters $R=0.1$ and $R=0.4$, respectively. The figure illustrates the effect of jet clustering on the event topology when transitioning from the parton to the hadron level. In particular, the jet-radius parameter influences the separation of the $b$-jets, which in turn affects the reconstruction of the invariant mass $M(b_1,b_2)$.
\begin{figure}[htb!]
    \centering
    \includegraphics[width=0.32\linewidth]{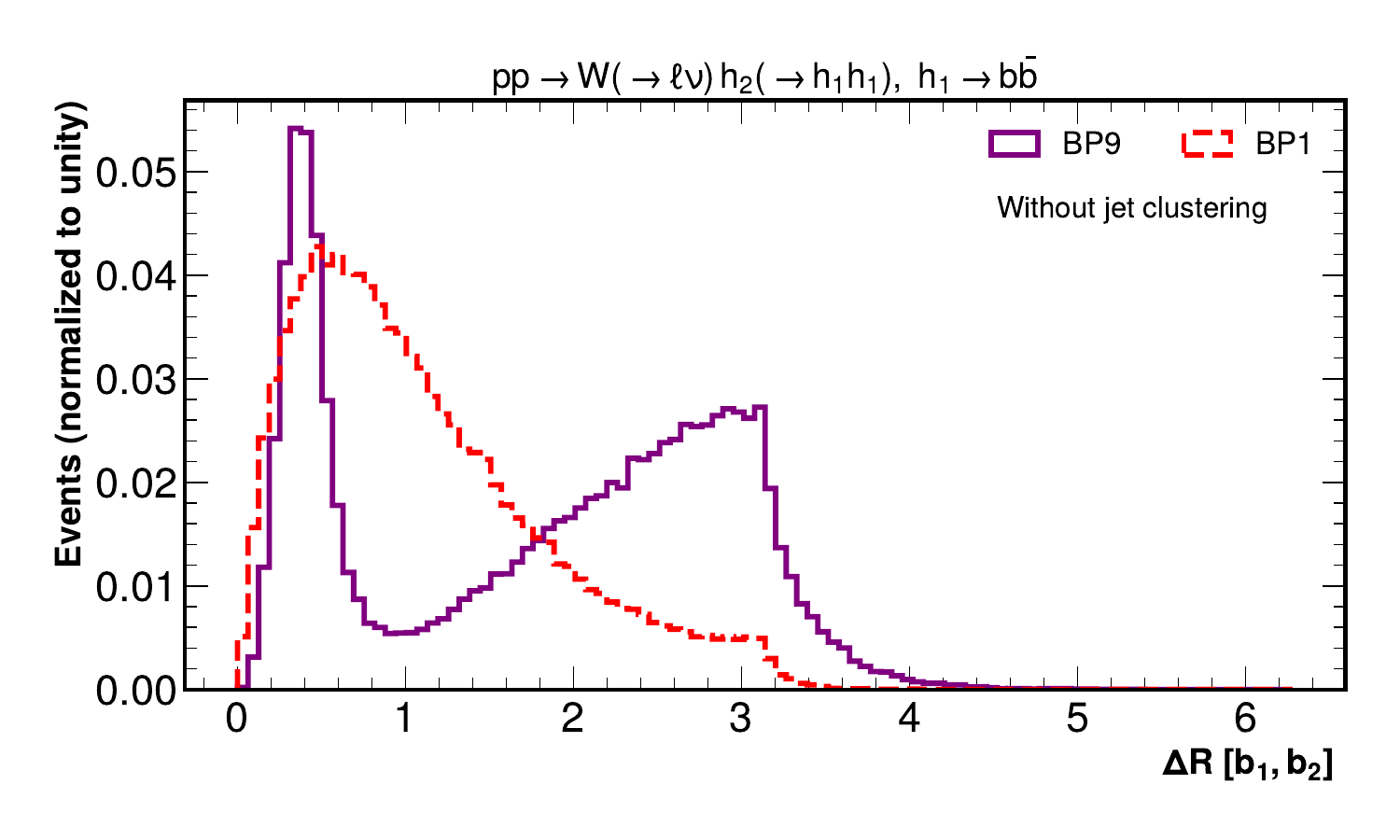}
    \includegraphics[width=0.32\linewidth]{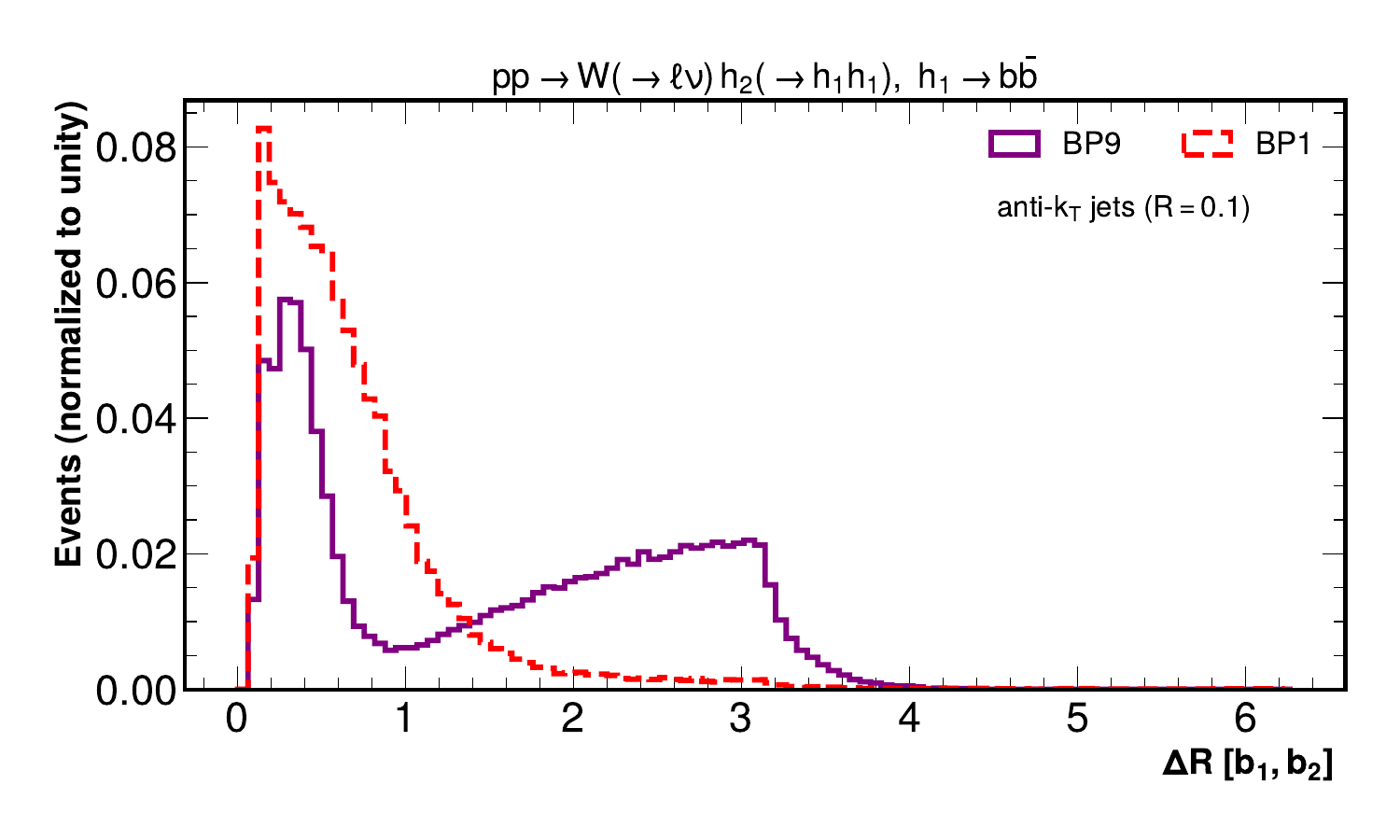}
    \includegraphics[width=0.32\linewidth]{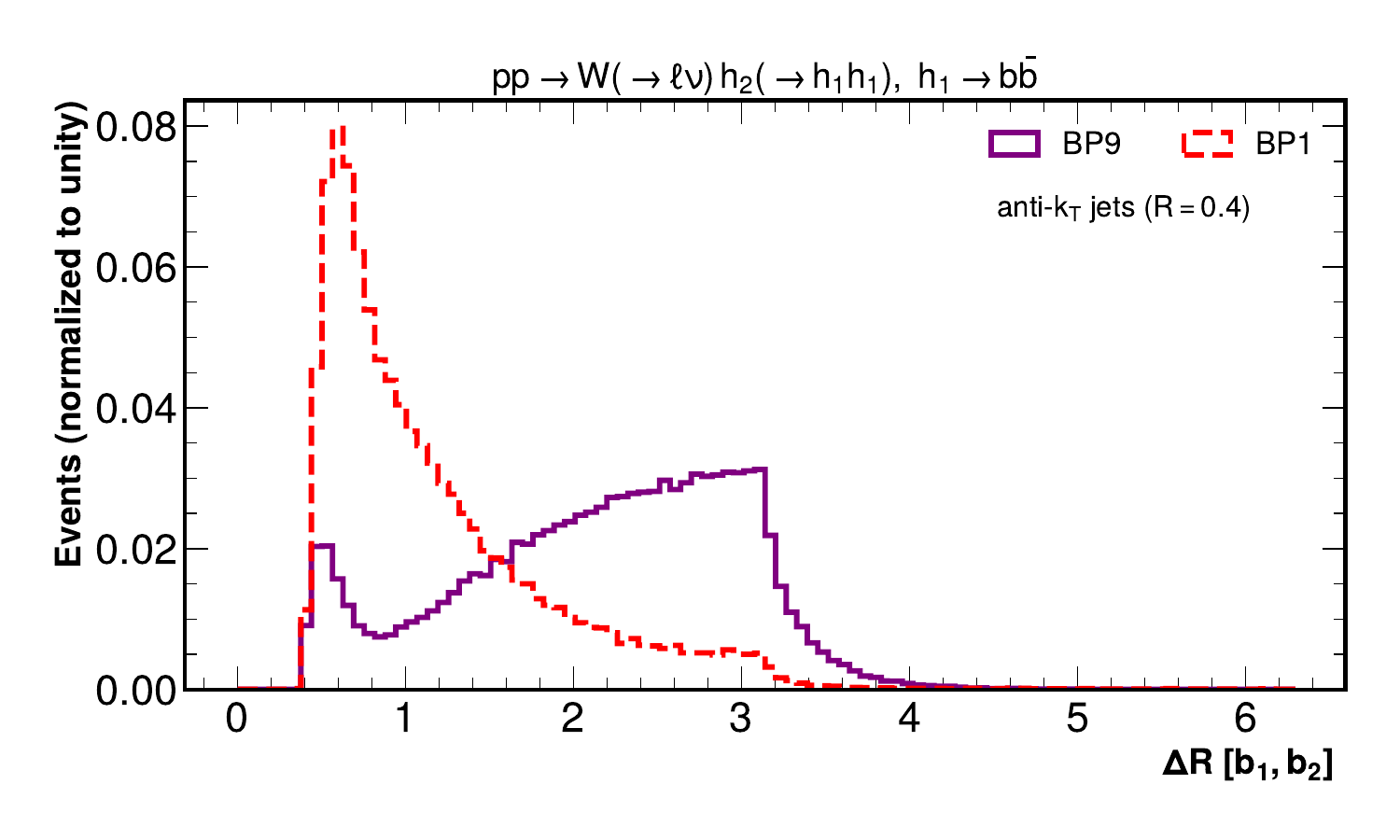}\\
    \includegraphics[width=0.32\linewidth]{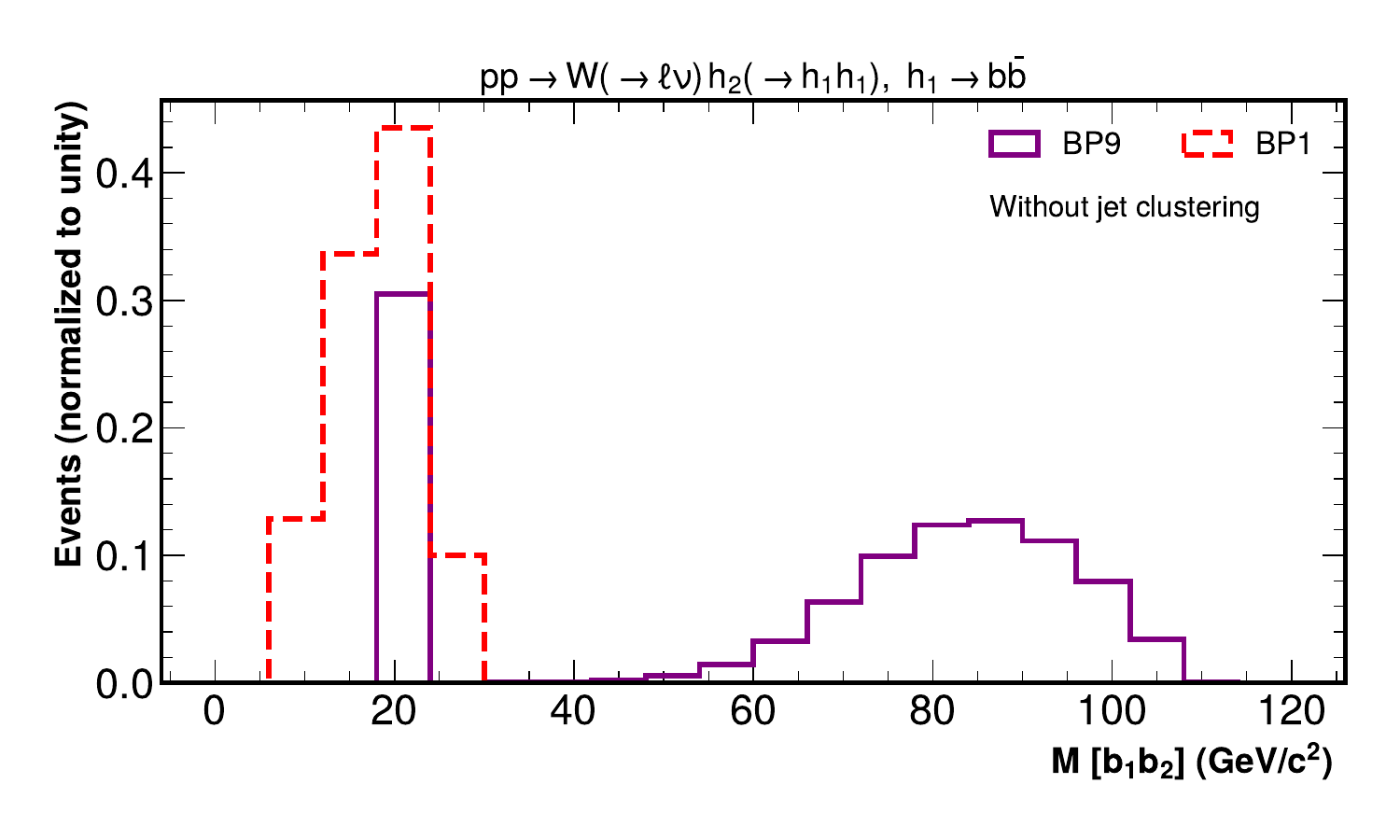}
    \includegraphics[width=0.32\linewidth]{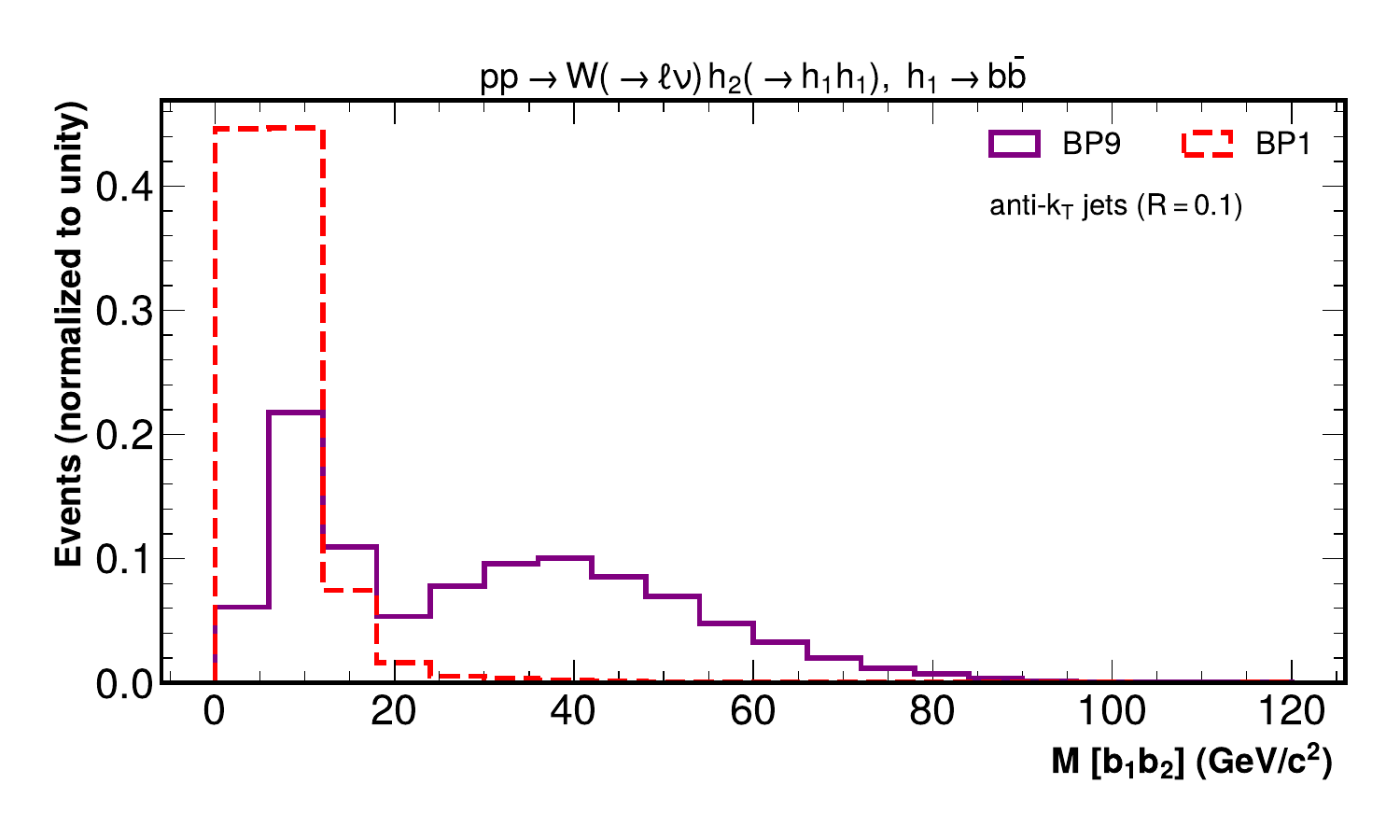}
    \includegraphics[width=0.32\linewidth]{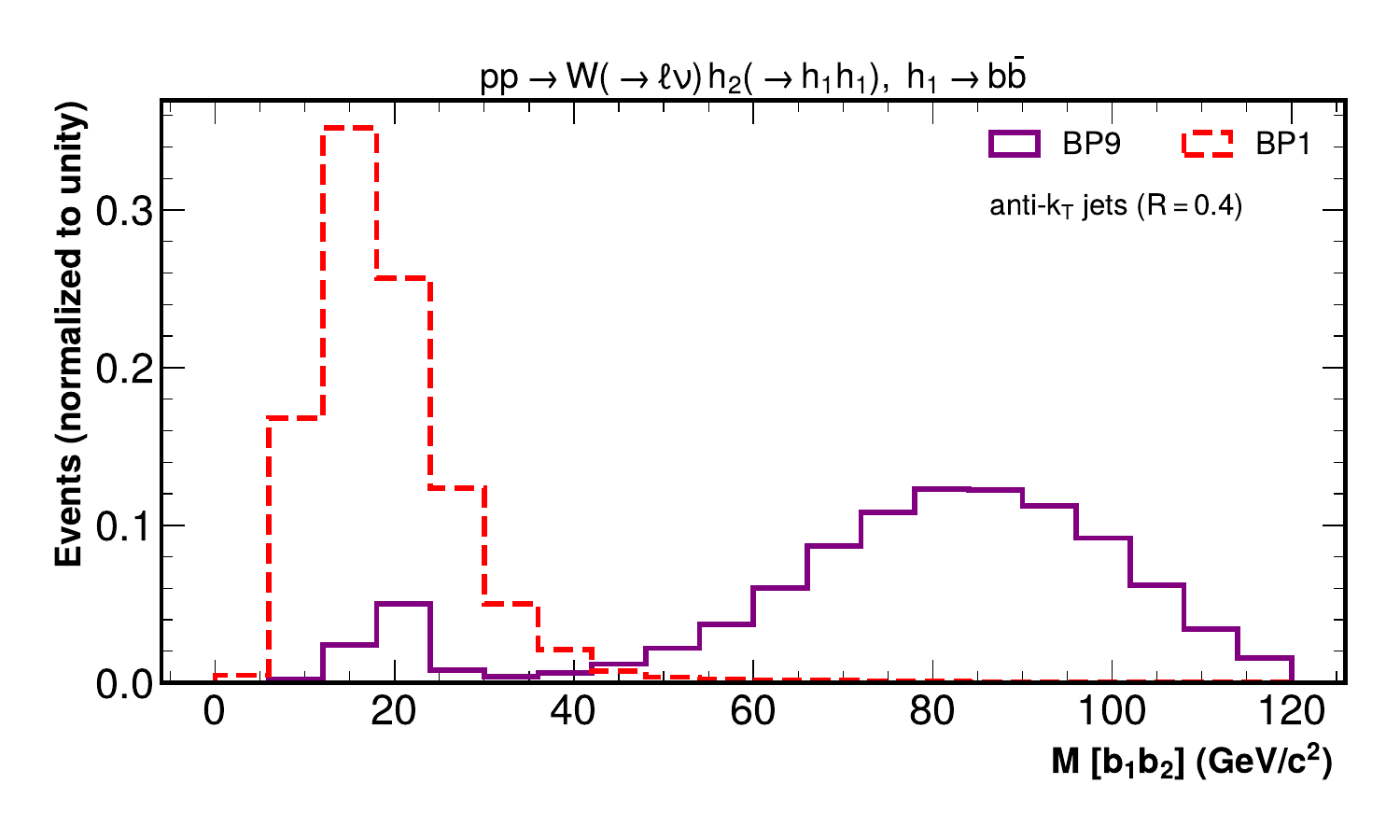}
\caption{Distributions of $\Delta R(b_1,b_2)$ {\sl (upper panels)} and the invariant mass $M(b_1,b_2)$ {\sl (lower panels)}. The left panels correspond to the parton level before jet clustering, whereas the middle and right panels show the hadron level after jet reconstruction with the anti-$k_T$ algorithm using radius parameters $R=0.1$ and $R=0.4$, respectively. 
}\label{fig:deltaR_Mb1b2}
\end{figure}

In general, the two leading $b$-jets, $b_1$ and $b_2$, are not necessarily the decay products of the same $h_1$ boson,
Therefore, the invariant mass
\[
M(b_1,b_2)
\]
does not always reconstruct the true $h_1$ mass and may exhibit a broader distribution or a shifted peak. This effect is more pronounced for benchmark points with larger $M_{2}$ values, where the boost of the intermediate $h_1$ states is significantly enhanced.

The benchmark point with $M_{2}\simeq 2M_{1}$ produces $h_1$ bosons nearly at rest, resulting in relatively soft but well-separated $b$-jets. In contrast, larger values of $M_{2}$ yield highly boosted $h_1$ bosons, whose decay products become increasingly collimated, affecting the reconstruction of the $b\bar b$ system.\\

For the evaluation of sensitivities, we use the Asimov significance expression for estimating the statistical significance given by \cite{Cowan:2010js}
\begin{align}
\label{eq:no-syst}
\text{Z}\lb S,B\rb\,=\,\sqrt{2 \left [(S + B) \cdot \ln\left(1 + S/B \right) - S \right]}.
\end{align}

In order to optimally reconstruct the light scalar $h_1$ decaying into a $b\bar{b}$ pair, we employ two complementary strategies depending on the event topology. For events containing at least four $b$-tagged jets, a full combinatorial reconstruction is performed using a $\chi^2$ minimization procedure. All possible pairings of the four leading $b$-tagged jets into two dijet systems are considered, and the combination that minimizes
\begin{equation}
\chi^2 = (M_{ij} - M_{1})^2 + (M_{kl} - M_{1})^2
\end{equation}
is selected, where $M_{ij}$ and $M_{kl}$ are the invariant masses of the two dijet candidates, and $M_{1}$ denotes the nominal mass of the scalar boson. This method exploits the presence of two $h_1 \to b\bar{b}$ decays in the signal and provides an efficient way to correctly assign the decay products in events with sufficiently large $b$-jet multiplicity.

However, the baseline event selection in this analysis targets final states with exactly one or two reconstructed leptons and at least 4-jet were at least two are identified as $b$-tagged jets, where events with only two $b$-jets dominate. In this regime, a full combinatorial reconstruction is not possible. Instead, we define a simplified observable based on the invariant mass of the available $b$-jet pair:
\begin{equation}
\Delta M_{bb}^{\mathrm{min}} = \min_{i<j} \left| M_{ij} - M_{1} \right|
\end{equation}
where the minimum runs over all possible pairs of $b$-tagged jets. For events with exactly two $b$-jets, this reduces to the absolute difference between the dijet invariant mass and the expected $h_1$ mass. This variable quantifies the compatibility of the reconstructed dijet system with the signal hypothesis and serves as a powerful discriminant against background processes.

In addition, the presence of a dilepton pair in the final state allows for further background suppression by requiring the invariant mass of the two leptons to lie within a window around the $Z$-boson mass. The dilepton invariant mass is defined as
\begin{equation}
M_{\ell\ell} = \sqrt{(p_{\ell_1} + p_{\ell_2})^2}
\end{equation}
where $p_{\ell_1}$ and $p_{\ell_2}$ are the four-momenta of the two leading leptons. Imposing a selection on $M_{\ell\ell}$ significantly reduces non-resonant backgrounds while preserving signal efficiency. The combination of the dilepton invariant mass requirement and the $\Delta M_{bb}^{\mathrm{min}}$ observable provides a robust and physically motivated strategy to discriminate signal from background, ensuring sensitivity across different kinematic regimes while maintaining a consistent interpretation of the reconstructed objects.

We apply a common set of selection cuts to suppress the SM backgrounds discussed in Section~\ref{sec:bgd} and enhance the signal significance for all benchmark points. The selection criteria adopted for both the single-lepton and dilepton channels are summarized in Table~\ref{tab:cuts}. Figure~\ref{fig:Z_vs_DeltaR} presents the signal significance, defined in Eq.~(\ref{eq:no-syst}), as a function of the $\Delta M_{bb}^{\rm min}$ cut. As can be observed, the optimal value of $\Delta M_{bb}^{\rm min}$ depends on the considered benchmark point. Nevertheless, an appropriate lower bound on $\Delta M_{bb}^{\rm min}$ can efficiently suppress the dominant backgrounds while retaining a sizable fraction of the signal events, thereby improving the sensitivity to the signal at both the LHC and the HL-LHC. In addition, Tables~\ref{cutflow_wp} and~\ref{cutflow_wm} present the cut-flow of the signal and background cross sections, given in \pb, for three representative benchmark points (BP 2,3 and 4) in the $4b\ell^+\nu_\ell$ and $4b\ell^-\bar{\nu}_\ell$ final states, respectively. The results demonstrate that the selected cuts significantly reduce the SM backgrounds while preserving an observable signal contribution. Table~\ref{tab:significance_combined} summarizes the expected statistical significances for the individual $W^+h_2$ and $W^-h_2$ channels, together with their combined significance, at integrated luminosities of $300$ and $3000~\mathrm{\fb}^{-1}$. \footnote{The selection efficiency is defined as $\epsilon = \sigma_{\mathrm{after\ cuts}}/\sigma_{\mathrm{before\ cuts}}$, where $\sigma_{\mathrm{before\ cuts}}$ and $\sigma_{\mathrm{after\ cuts}}$ denote the production cross sections before and after the event selection, respectively.}
\begin{table}[htb!]
	\centering
	\setlength{\tabcolsep}{6pt}
	\renewcommand{\arraystretch}{1.3}
	\begin{tabular}{l|c|c|c|c|c|c}
		\toprule
		\textbf{Channel} & \textbf{Cut 1} & \textbf{Cut 2} & \textbf{Cut 3} & \textbf{Cut 4} & \textbf{Cut 5} & \textbf{Cut 6} \\
		\midrule
		$W h_2$ & $N_b \ge 2$ & $N_\ell = 1$ & $N_j = 2$ & $\Delta R_{bb} \in [0.4,4.0]$ & $\Delta M_{bb}^{\text{min}}$ & -- \\
		$Z h_2$ & $N_b \ge 2$ & $N_\ell = 2$ & $N_j = 2$ & $\Delta R_{bb} \in [0.4,4.0]$ & $\Delta M_{bb}^{\text{min}}$ & $M_{\ell\ell} \in [80,100]$ \\
		\bottomrule
	\end{tabular}
	\caption{Selection cuts optimized for signal significance over background. The same baseline cuts are applied for all benchmark points, with the dilepton channel requiring an additional invariant mass cut on the lepton pair.}
	\label{tab:cuts}
\end{table}

\begin{figure}
    \centering
    \includegraphics[width=0.45\linewidth]{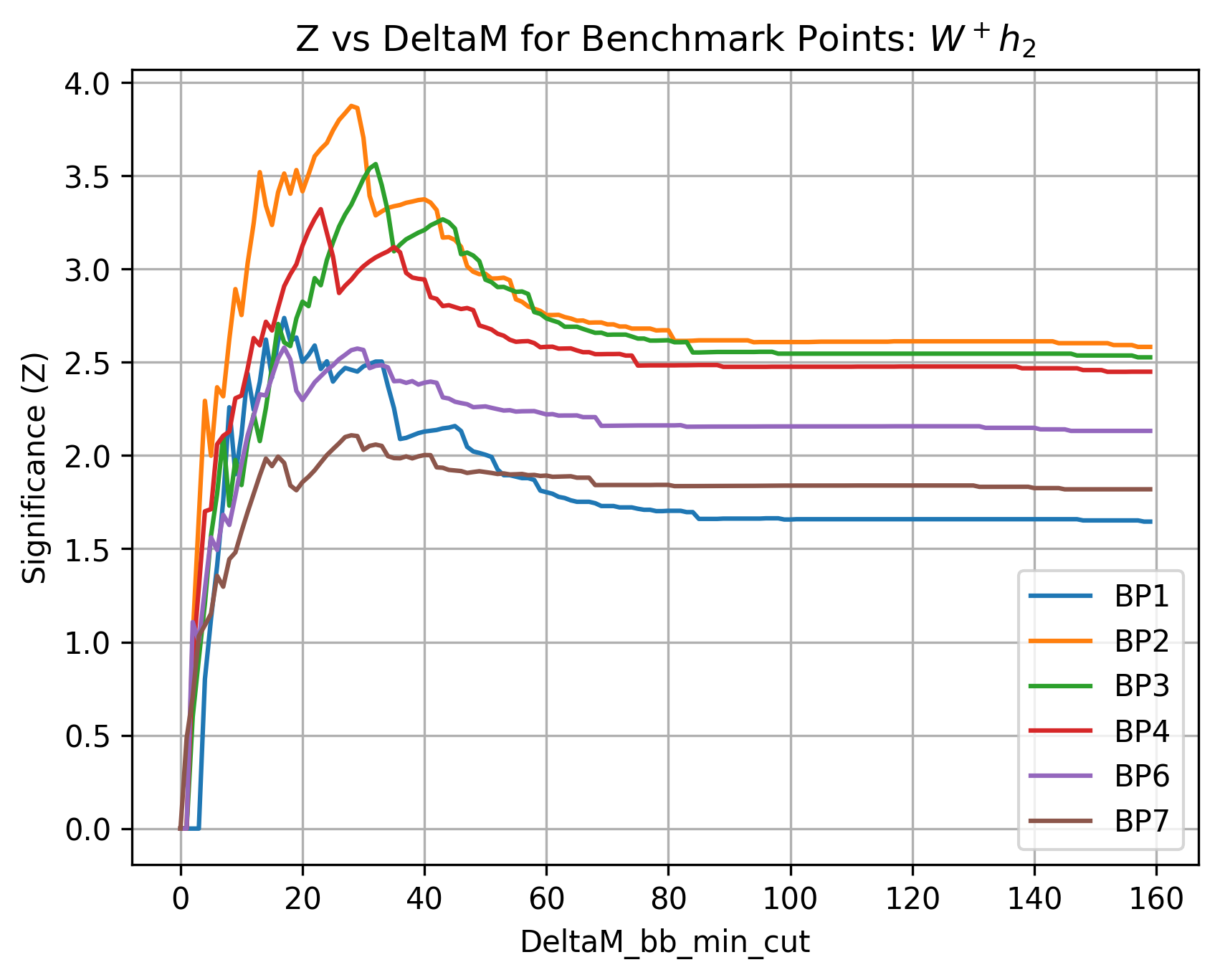}
    \includegraphics[width=0.45\linewidth]{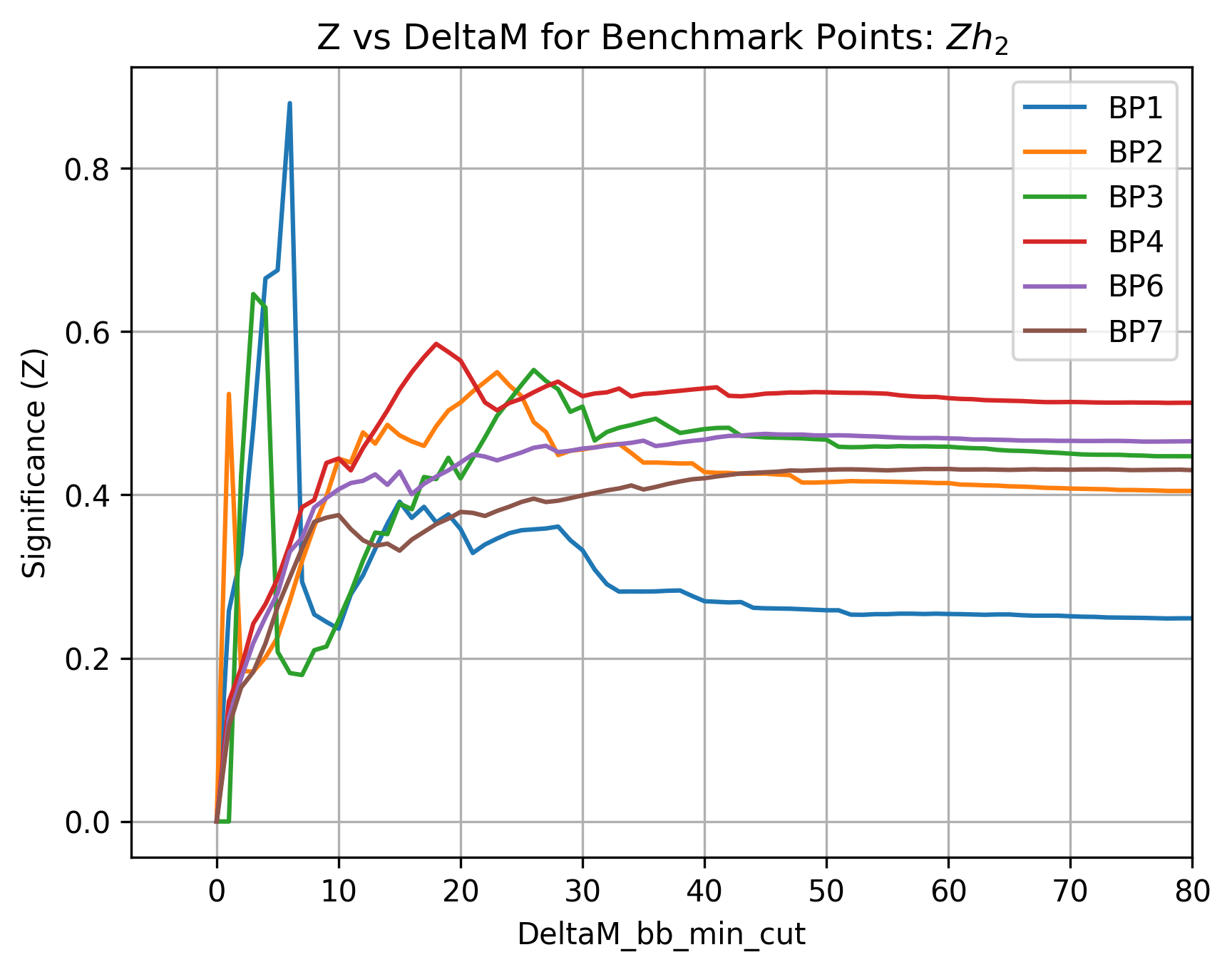}
\caption{Signal significance $Z$ as a function of the invariant-mass window cut $\Delta M_{bb}$ for the benchmark points BP1--BP7, after applying Cuts~1--4 of Table~\ref{tab:cuts}. The {\sl left (right)} panel corresponds to the $W^+h_2$ ($Zh_2$) production channel. The peak of each curve indicates the optimal $\Delta M_{bb}$ selection, balancing signal efficiency and background suppression.
}    \label{fig:Z_vs_DeltaR}
\end{figure}

\begin{table*}[htbp!]
\centering %
\setlength{\tabcolsep}{8pt}
\renewcommand{\arraystretch}{1.2}
\begin{adjustbox}{max width=\textwidth}	
	\begin{tabular}{lcccccc}
		\toprule
		\multirow{2}{*}{Cuts} & \multicolumn{2}{c}{BP2} & \multicolumn{2}{c}{BP3} & \multicolumn{2}{c}{BP4} \\
		\cmidrule(lr){2-3} \cmidrule(lr){4-5} \cmidrule(lr){6-7}
		& Signal & Background & Signal & Background & Signal & Background \\
		\midrule
		Basic & 0.06299 & 565.08 & 0.05661 & 565.08 & 0.03672 & 565.08 \\
		\midrule
		Cut1 & 0.003160   &   32.26 & 0.003344  & 32.26 & 0.003520  & 32.26 \\
		Cut2 & 0.001801   &   17.63 & 0.001901  & 17.63 & 0.002069  &   17.63 \\
		Cut3 & 0.000872   &   0.1188 & 0.000894 & 0.1188 & 0.000900 &  0.1188 \\
		Cut4 & 0.000872   &   0.1172 & 0.000894  & 0.1172 & 0.000900  & 0.029550 \\
		Cut5 & 0.000760   &   0.01114 & 0.000721 &0.011143 & 0.000685   &   0.011143 \\
		\midrule
		Eff. [\%] & 1.20 & 0.00197 & 1.27 & 0.00197 & 1.86 & 0.00197 \\
		\bottomrule
	\end{tabular}
\end{adjustbox}
\caption{Cut-flow for the $W^{+}h_2$ signal and the relevant background processes at $\sqrt{s}=13.6$~\TeV. The cross sections, given in \pb, are reported after each successive selection cut present in Tab~\ref{tab:cuts}. 
}\label{cutflow_wp}
\end{table*}

\begin{table*}[htbp!]
\centering %
\setlength{\tabcolsep}{8pt}
\renewcommand{\arraystretch}{1.2}
\begin{adjustbox}{max width=\textwidth}	
	\begin{tabular}{lcccccc}
		\toprule
		\multirow{2}{*}{Cuts} & \multicolumn{2}{c}{BP2} & \multicolumn{2}{c}{BP3} & \multicolumn{2}{c}{BP4} \\
		\cmidrule(lr){2-3} \cmidrule(lr){4-5} \cmidrule(lr){6-7}
		& Signal & Background & Signal & Background & Signal & Background \\
		\midrule
		Basic & 0.04362 & 413.58 & 0.03907 & 413.58 & 0.02502 & 413.58 \\
		\midrule
		Cut1 &  0.002311  &  32.13  & 0.002548 & 32.13 & 0.002568  & 32.13 \\
		Cut2 &  0.001233  &  17.66  & 0.001353  & 17.66 & 0.001410  & 17.66  \\
		Cut3 &  0.000612  &  0.102375  & 0.000627  & 0.102375 & 0.000626  & 0.102375  \\
		Cut4 &  0.000612  &  0.101693  & 0.000627  & 0.101693 & 0.000626  & 0.101693  \\
		Cut5 &  0.000532  &  0.011973  & 0.000511  & 0.011973 & 0.000497  & 0.012806   \\
		\midrule
		Eff. [\%] & 1.22 & 0.002895 & 1.30 & 0.002895 & 1.98 &  0.00309\\
		\bottomrule
	\end{tabular}
\end{adjustbox}
\caption{Cut-flow for the $W^{-}h_2$ signal and the relevant background processes at $\sqrt{s}=13.6$~\TeV. The cross sections, given in \pb, are reported after each successive selection cut present in Tab~\ref{tab:cuts}. }\label{cutflow_wm}
\end{table*}

\begin{table}[htpb!]
	\centering
	\setlength{\tabcolsep}{4pt}
	\renewcommand{\arraystretch}{1.2}
	\begin{tabular}{lccccccccccc}
		\toprule
		\multirow{2}{*}{$\mathcal{L}$ [\fb$^{-1}$]} & \multirow{2}{*}{Channel} & \multicolumn{9}{c}{Benchmark Points} \\
		\cmidrule(lr){3-11}
		& & BP1 & BP2 & BP3 & BP4 & BP5 & BP6 & BP7 & BP8 & BP9 \\
		\midrule
		\multirow{3}{*}{300} & $W^+h_2$ & 2.92 & 3.90 & 3.70 & 3.52 & 2.26 & 2.54 & 2.23 & 1.41 & 1.01 \\
		& $W^-h_2$ & 2.05 & 2.64 & 2.54 & 2.39 & 1.52 & 1.88 & 1.52 & 1.06 & 0.74 \\
		& Combined &  3.48 & 4.61 & 4.40 & 4.15 & 2.66 & 3.16 & 2.70 & 1.75 & 1.25 \\
		\midrule
		\multirow{3}{*}{3000} & $W^+h_2$ & 9.22 & 12.33 & 11.70 & 11.12 & 7.16 & 8.02 & 7.06 & 4.45 & 3.21 \\
		& $W^-h_2$ & 6.47 & 8.36 & 8.03 & 7.56 & 4.81 & 5.95 & 4.81 & 3.35 & 2.35 \\
		& Combined & 10.99 & 14.59 & 13.91 & 13.12 & 8.42 & 9.98 & 8.53 & 5.54 & 3.96 \\
		\bottomrule
	\end{tabular}
	\caption{Discovery significances for the $W^+h_2$, $W^-h_2$, and combined $W^+h_2+W^-h_2$ channels at $\sqrt{s}=13.6$~\TeV for integrated luminosities of $\mathcal{L}=300$ and $3000$~\fb$^{-1}$.}
	\label{tab:significance_combined}
\end{table}

We also investigated the $Z\,h_2$ channel with a dilepton decay of the $Z$ boson. Unfortunately for LHC-like integrated luminosities we did not find a set of cuts leading to satisfactory significances. For HL-LHC integrated luminosities, significances stay below 3. We therefore do not further comment on this channel in the rest of this work. 

 Figure~\ref{fig:signi} shows the statistical significance in the $(M_2-2M_1,\,M_2)$ plane for the $W^+h_2$ {\sl (upper panels)}, $W^-h_2$ {\sl (middle panels)}, and the combination~\footnote{The combined significance is obtained by summing the signal and background event yields from the $W^+h_2$ and $W^-h_2$ channels after the full event selection and subsequently evaluating the statistical significance using Eq.~(\ref{eq:no-syst}).} of both channels {\sl (lower panels)}. For the $W^+h_2$ channel, evidence for the signal ($Z \geq 3$) is achieved for three benchmark points (BP2, BP3, and BP4) out of the nine scenarios considered, already at an integrated luminosity of $300~\mathrm{fb}^{-1}$. In contrast, the $W^-h_2$ channel exhibits lower sensitivities and remains below the evidence threshold for all benchmark points under consideration. The higher sensitivity of the $W^+h_2$ channel originates from the larger $u$-quark parton luminosity in the proton, which enhances the production rate relative to $W^-h_2$.

The highest significances are obtained for BP2--BP4, corresponding to intermediate values of $M_2$, where the production cross sections remain sizeable and the resulting $b$-jets are sufficiently separated to be efficiently reconstructed. Furthermore, the presence of a clean leptonic signature provides strong discrimination against SM backgrounds, thereby enhancing the overall signal sensitivity.

For benchmark points with larger values of $M_2$, the sensitivity gradually decreases, primarily due to the reduction in the production cross section and the resulting loss of signal statistics. This behaviour is particularly evident for BP8 and BP9, which exhibit significantly lower sensitivities than the intermediate-mass scenarios. Similarly, benchmark points with very light scalar masses, $M_1 \lesssim 20$~\GeV, are more challenging, as the decay products tend to be either too soft or highly collimated. Consequently, the efficiency for reconstructing the two $b$-jets as separate resolved objects is reduced, leading to a loss of sensitivity within the resolved-jet framework adopted in this analysis. For this reason, the present study is restricted to BPs with $M_1 > 20$~\GeV. The combination of the $W^+h_2$ and $W^-h_2$ channels leads to an overall improvement in the expected signal significance. In particular, BP1 and BP6, which individually remain below the evidence threshold, reach evidence-level sensitivity ($Z \geq 3$) after combining both channels.

The extrapolation to the HL-LHC significantly extends the discovery reach. Several benchmark points that provide only evidence-level or marginal sensitivity at $300~\mathrm{fb}^{-1}$ surpass the discovery threshold at $3000~\mathrm{fb}^{-1}$, highlighting the crucial role of increased luminosity. Nevertheless, BPs with large mass splittings, $M_2-2M_1$ (BP8 and BP9), remain challenging even at the HL-LHC within the resolved-jet analysis presented here. Such regions could benefit from advanced reconstruction techniques, including jet-substructure methods and improved $p_T$ $b$-tagging algorithms, which are expected to enhance the signal sensitivity. A detailed study of these techniques is beyond the scope of the present work. 

Combining the $W^+h_2$ and $W^-h_2$ channels at the HL-LHC significantly enhances the expected signal significance. As a result, all benchmark points except BP9 exceed the discovery threshold ($Z>5$).

\begin{figure}[htb!]
    \centering
    \includegraphics[width=0.45\linewidth]{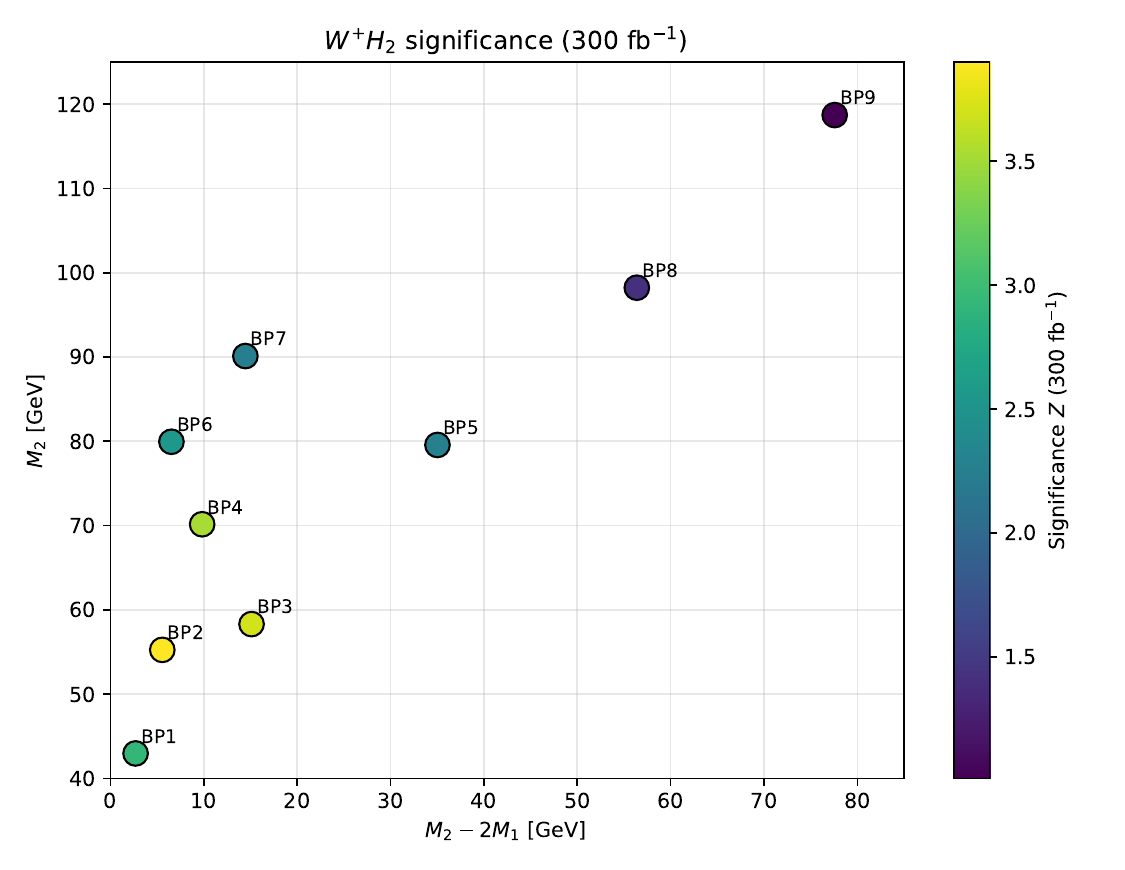}
    \includegraphics[width=0.45\linewidth]{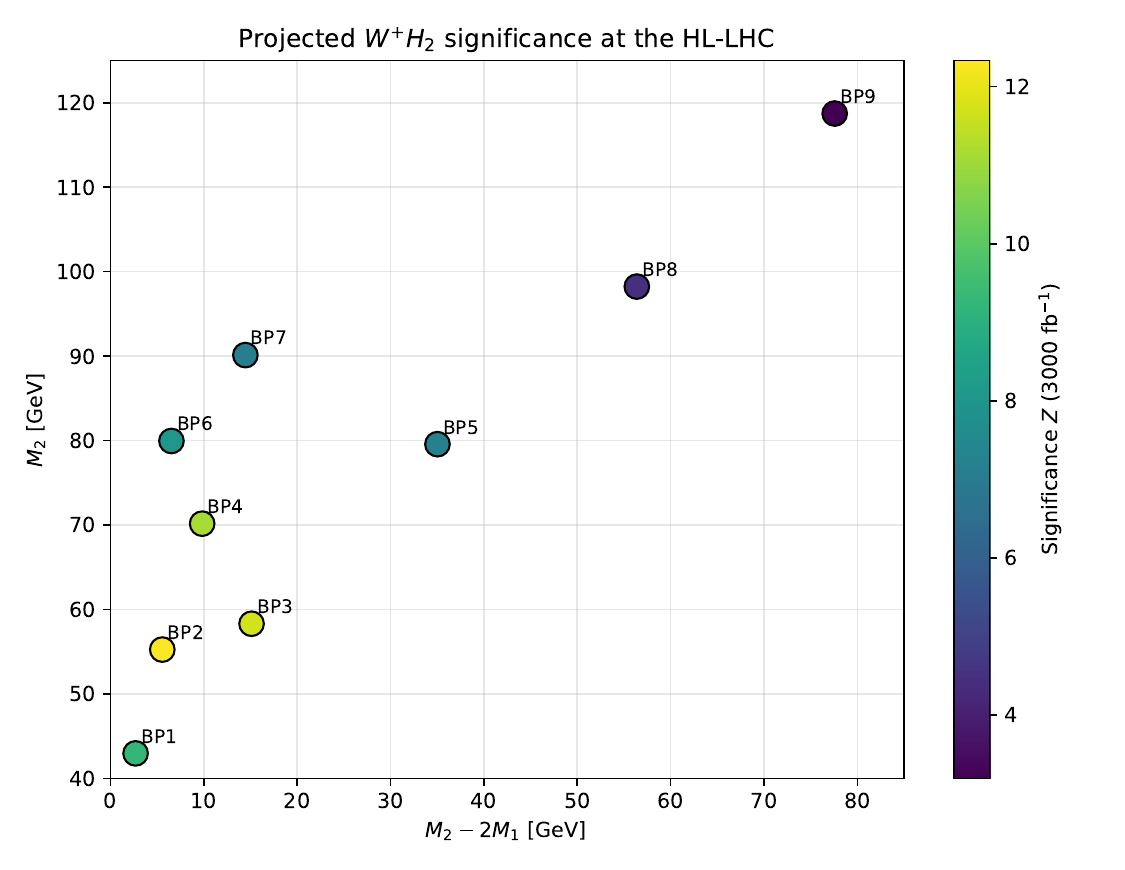}\\
    \includegraphics[width=0.45\linewidth]{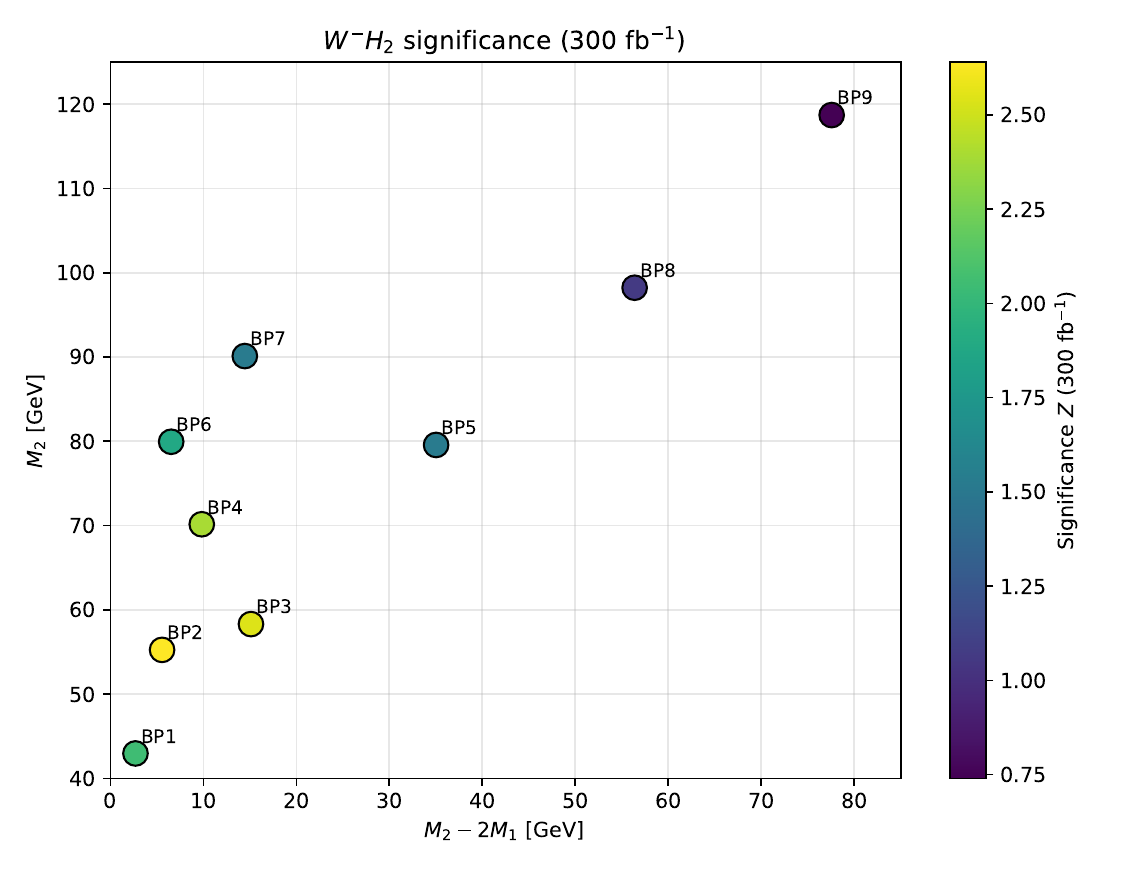}
    \includegraphics[width=0.45\linewidth]{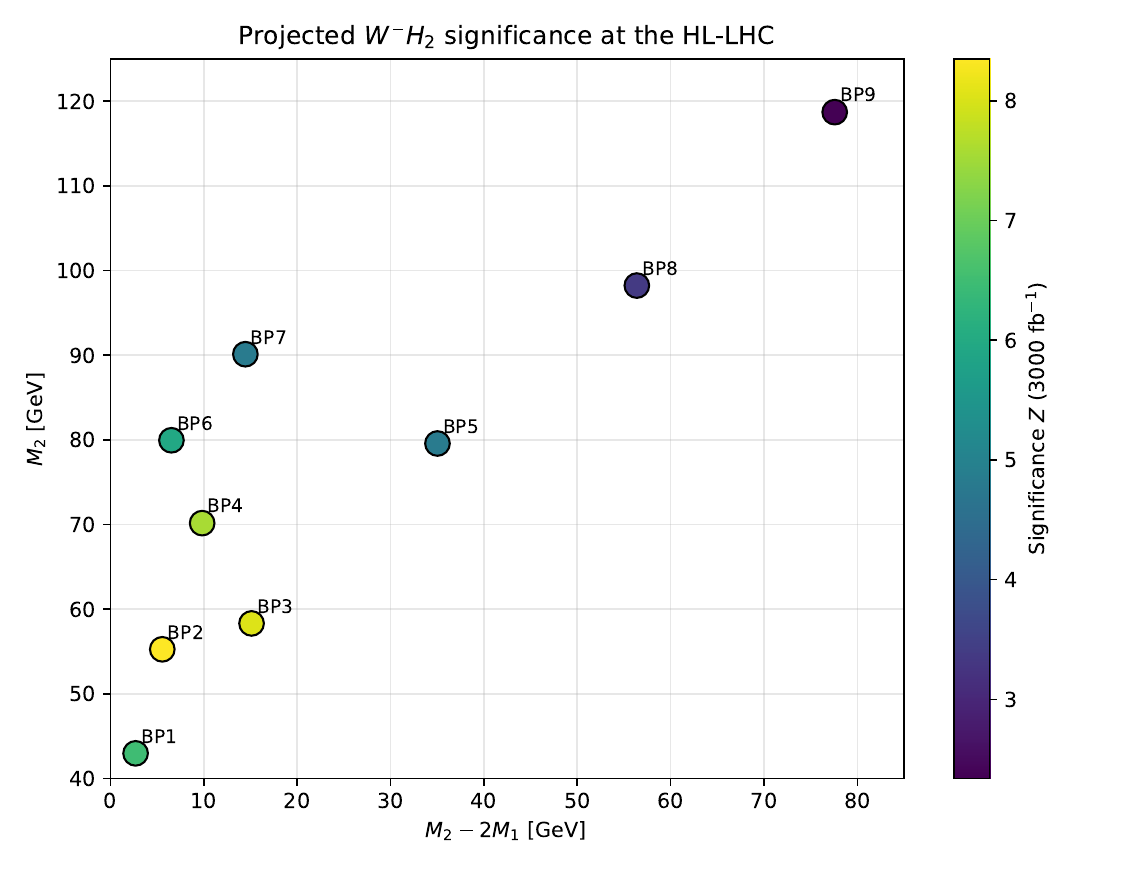}
    \includegraphics[width=0.45\linewidth]{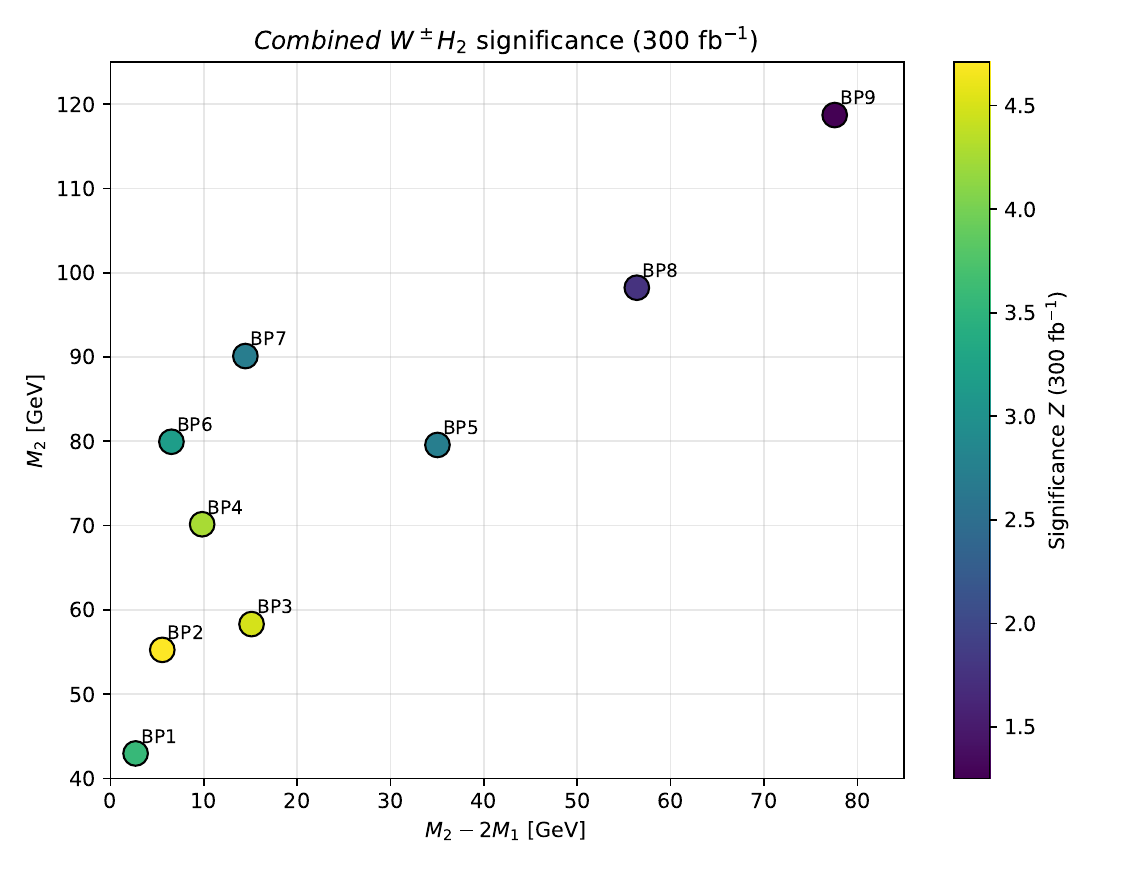}
    \includegraphics[width=0.45\linewidth]{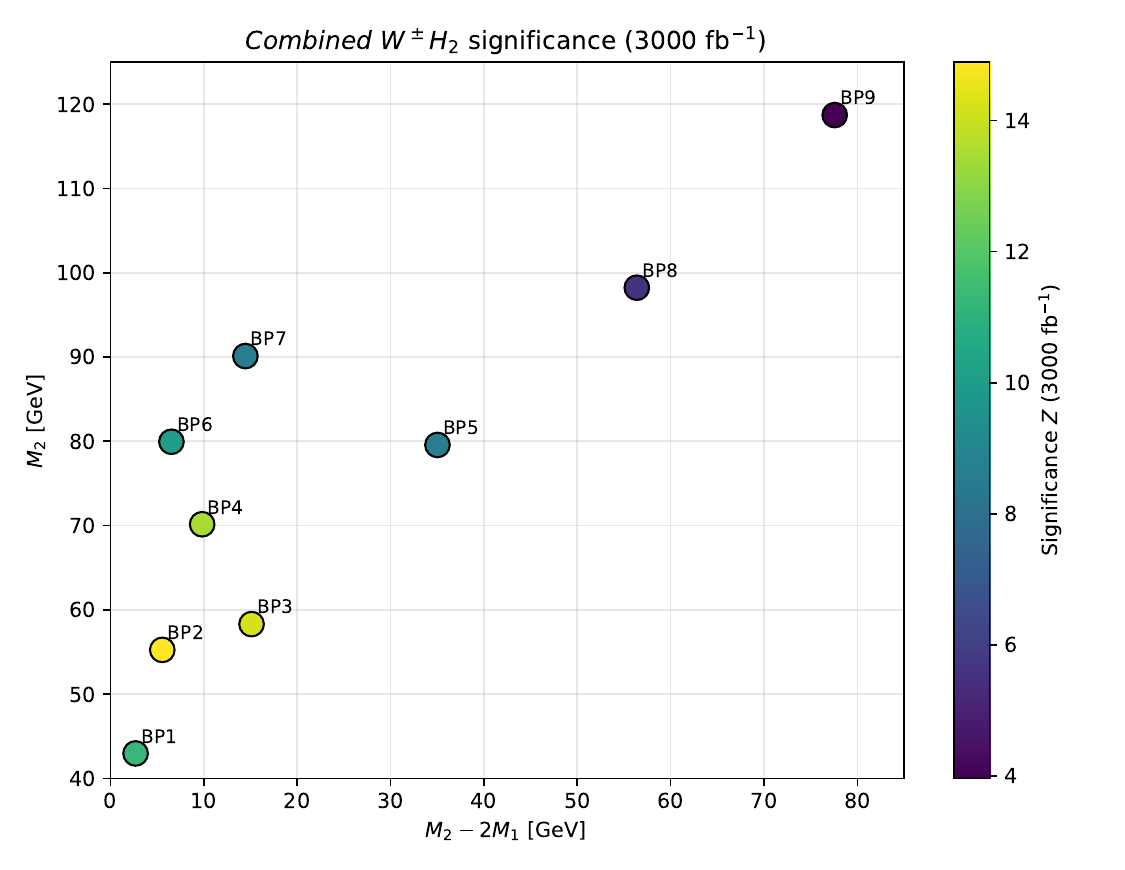}
\caption{Expected statistical significance in the $(M_2-2M_1,\,M_2)$ plane for the $W^+h_2$ {\sl (upper panels)}, $W^-h_2$ {\sl (middle panels)}, and the combination of both channels {\sl (lower panels)}. The left panels correspond to an integrated luminosity of $300~\mathrm{fb}^{-1}$ (Run~3), while the right panels show the HL-LHC projection for $3000~\mathrm{fb}^{-1}$. The color scale indicates the statistical significance, $Z$.}
\label{fig:signi}
\end{figure}

Finally, we briefly want to comment on the effect of including higher-order corrections. As discussed above, we here perform a study at leading-oder for signal and most background processes. While a full NLO study is beyond the scope of the current work, we briefly estimate how our significances would change if constant $K$-factors would be applied to both signal and background processes. The corresponding discussion can be found in Appendix \ref{sec:K-Factor}.
\section{Conclusions}
\label{sec:conclusions}

We have studied the prospects for probing an extended Higgs sector featuring a heavy scalar $h_2$ that decays into a pair of lighter scalars, $h_2 \to h_1 h_1 \to b\bar b b\bar b$, in association with a vector boson at the LHC. The analysis targets final states with one or two isolated leptons originating from the leptonic decay of the associated $W$ or $Z$ boson, accompanied by multiple $b$-tagged jets. A standard resolved-jet strategy with conventional object selections has been employed.

Our results demonstrate that associated production channels provide a powerful probe of this class of scenarios over a wide region of the $(M_{1}, M_{2})$ parameter space. For several benchmark points,  evidence-level sensitivity can already be achieved with an integrated luminosity of $300~\mathrm{fb}^{-1}$, driven primarily by the $W^+ h_2$ production modes.  The projected reach at the HL-LHC significantly extends the coverage, with most benchmark scenarios yielding high statistical significance at $3000~\mathrm{fb}^{-1}$. The sensitivity is found to be maximal for intermediate values of $M_{2}$, where the production cross section remains sizable and the decay products can be efficiently reconstructed using resolved jets.

In contrast, scenarios with very light $h_1$ states or large $h_2$ masses remain challenging within the present analysis framework, owing to reduced signal rates and softer or more collimated $b$-jet kinematics. These regions could benefit from dedicated reconstruction techniques, such as jet substructure methods or improved low-$p_T$ $b$-tagging algorithms, which are beyond the scope of this work.

Overall, this study highlights the strong potential of associated Higgs production with leptonic vector-boson decays as a robust and complementary avenue for exploring non-standard Higgs sectors at the LHC and the HL-LHC.

\section*{Acknowledgments}
TR and MO are supported by the Croatian Science Foundation (HRZZ) under Grant No. HRZZ-IP-2022-10-2520. They also acknowledge support from the Deutsche Forschungsgemeinschaft (DFG, German Research Foundation) under Germany’s Excellence Strategy — EXC~2121 “Quantum Universe” — Project No. 390833306. TR and MO further thank the CERN Theory Group for their hospitality during the completion of part of this work. MO also acknowledges the hospitality of the ATLAS High Energy Physics Group at the Department of Physics and Astronomy, University of Sheffield, UK, where part of this research was carried out. In addition, MO acknowledges support from COMETA (COmprehensive Multiboson Experiment–Theory Action) through a Short-Term Scientific Mission (STSM) within COST Action CA22130. PS is supported by Basic Science Research Program through the National Research Foundation of Korea (NRF) funded by the Ministry of Education through the Center for Quantum Spacetime (CQUeST) of Sogang University with grant number RS-2020-NR049598 and by the Ministry of Science and ICT with grant number RS-2025-24523022.

\section*{Appendix}
\appendix

\renewcommand{\thetable}{\Alph{section}.\arabic{table}}
\section{Decay and total widths}

\setcounter{table}{0}

\label{withs-MG5}
\begin{table}[H]
    \centering
    \begin{tabular}{c|cccc}
         BPs&BR($h_2\to h_1 h_1$)\ & BR($h_1\to b\bar{b}$) \ & $\Gamma_{h_1}(10^{-3}\,\MeV)$\ & $\Gamma_{h_2}(\MeV)$ \\\hline
         BP1&$0.82$&$0.94$&$3.5$&$0.49$ \\
         BP2&$0.89$&$0.94$&$4.9$&$1.1$ \\
         BP3&$0.92$&$0.94$&$3.9$&$1.7$ \\
         BP4&$0.94$&$0.95$&$6.4$&$2.5$ \\
         BP5&$0.96$&$0.94$&$4.1$&$4.1$ \\
         BP6&$0.94$&$0.95$&$8.1$&$3.1$ \\
         BP7&$0.96$&$0.95$&$8.4$&$5.3$ \\
         BP8&$0.97$&$0.94$&$3.7$&$7.4$ \\
         BP9&$0.98$&$0.94$&$3.6$&$12.7$ \\\hline
    \end{tabular}
\caption{Decay widths and branching ratios used in the event simulation, calculated with \texttt{MadGraph5\_aMC@NLO}.}    \label{tab:MG5-withsBR}
\end{table}

\section{Irreducible background}
\setcounter{table}{0}
\label{ir_BG}

To assess the impact of non-resonant contributions on the signal, we compare the resonant production process
\[
pp \to h_2W^+ \to h_1h_1W^+ \to 4b\,\ell^+\nu_\ell,
\]
with the complete partonic process
\[
pp \to b\bar{b}b\bar{b}\,\ell^+\nu_\ell,
\]
which includes all resonant and non-resonant Standard Model amplitudes as well as new physics contributions, together with their interference. The non-resonant contribution is dominated by the continuum production process
\[
pp\to W^+gg,\qquad g\to b\bar b,
\]
which constitutes the largest irreducible background to the signal.

The resonant signal contribution is generated in \texttt{MadGraph5\_aMC@NLO} using

\begin{lstlisting}[language=bash,
                   basicstyle=\ttfamily\small,
                   frame=single,
                   breaklines=true]
generate p p > h h w+, h > b b~, h > b b~, w+ > l+ vl QCD=10 QED=10
\end{lstlisting}

where the process is restricted to the resonant $W^+h_2$ topology followed by the decays
$h_2\to h_1h_1$, $h_1\to b\bar b$, and
$W^+\to\ell^+\nu_\ell$.

For comparison, the complete final state is generated using

\begin{lstlisting}[language=bash,
                   basicstyle=\ttfamily\small,
                   frame=single,
                   breaklines=true]
generate p p > b b~ b b~ l+ vl QCD=10 QED=10
\end{lstlisting}

which includes all Feynman diagrams contributing to
$pp\to b\bar b b\bar b\ell^+\nu_\ell$, including both resonant and non-resonant contributions as well as their interference.

To isolate the dominant non-resonant component, we additionally generate

\begin{lstlisting}[language=bash,
                   basicstyle=\ttfamily\small,
                   frame=single,
                   breaklines=true]
generate p p > w+ g g, g > b b~, w+ > l+ vl QCD=10 QED=10
\end{lstlisting}

which contains the dominant continuum production mechanism with an on-shell $W^+$ boson and allows a direct comparison with the resonant signal topology.

The corresponding production cross sections at $\sqrt{s}=13.6$~\TeV are summarized in Table~\ref{tab:inter} before and after the baseline and analysis selections. For completeness, Tables~\ref{tab:xs_after_cuts} and~\ref{tab:xs_after_cuts-wm} list the cross sections of all background processes after the full event selection for each benchmark point.

Before the event selection, the complete process receives comparable contributions from both the resonant and non-resonant amplitudes, leading to production cross sections of similar magnitude. After applying the baseline and analysis selections, however, the resonant contribution becomes significantly enhanced relative to the continuum background. For BP1--BP4, the resonant signal is comparable to or even exceeds the dominant non-resonant contribution, whereas its relative importance decreases for larger scalar masses owing to the reduction in the production cross section.

For BP1, the resonant signal slightly exceeds the cross section of the complete process after the event selection, indicating a small destructive interference between the resonant and non-resonant amplitudes. Nevertheless, the difference remains at the percent level and has a negligible impact on the final signal yield. Overall, the comparison demonstrates that the non-resonant contribution, including its interference with the resonant amplitude, has only a limited effect on the signal region after the event selection. Consequently, the resonant process alone provides an excellent approximation for the remainder of this analysis.
 
\begin{table}[htb!]
    \centering
    \setlength{\tabcolsep}{6pt}
    \renewcommand{\arraystretch}{1.2}
    \begin{tabular}{c|cccc}
        \toprule
        BPs & BP1 & BP4 & BP7 & BP9 \\
        \midrule
        $M_{1}$ [\GeV] & 20.13 & 30.16 & 37.83 & 20.58 \\
        $M_{2}$ [\GeV] & 42.96 & 70.16 & 90.12 & 118.72 \\
        \midrule
        $pp \to hhW^+ \to bbbb\,\ell^+\nu_\ell$(signal) & $0.10002\,(7)$ & $0.03672\,(3)$ & $0.01918\,(1)$ & $0.008427\,(6)$ \\
        $pp \to bbbb\,\ell^+\nu_\ell$(full process)  & $0.1661\,(6)$ & $0.1139\,(3)$ & $0.1022\,(2)$ & $0.0943\,(3)$ \\
       $pp \to W^+gg,\; g \to b\bar{b},\; W^+ \to \ell^+\nu_\ell$ & $0.09588\,(6)$ & $0.09588\,(6)$ & $0.09588\,(6)$ & $0.09588\,(6)$ \\

        \midrule
        \multicolumn{5}{c}{After cuts in Eq.~(\ref{based_cuts1}--\ref{based_cuts2}) and Table~\ref{tab:cuts}} \\
        \midrule
        $pp \to hhW^+ \to bbbb\,\ell^+\nu_\ell$(signal) & $0.000394$ & $0.000685$ & $0.000611$ & $0.000532$ \\
        $pp \to bbbb\,\ell^+\nu_\ell$(full process) & $0.000365$ & $0.001048$ & $0.001186$ & $0.00228$ \\
        $pp \to W^+gg \to bbbb\,\ell^+\nu_\ell$ & $0.000164$ & $0.000377$ & $0.000639$ & $0.00156$ \\
        \bottomrule
    \end{tabular}
    \caption{Signal cross sections in \pb at $\sqrt{s}=13.6$~\TeV for the signal process $pp \to hhW^+ \to 4b\,\ell^+\nu_\ell$, and for the full signal process including all contributions. Values in parentheses indicate integration errors.}
    \label{tab:inter}
\end{table}

\begin{table}[htb!]
	\centering
	\setlength{\tabcolsep}{4pt}
	\renewcommand{\arraystretch}{1.2}
	\begin{tabular}{lcccccccc}
		\toprule
		\multirow{2}{*}{BP} & \multicolumn{8}{c}{Cross-section [\fb] after selection cuts} \\
		\cmidrule(lr){2-9}
		&$Xs(BP_i)$& $t\bar{t}$+jets & $W$+jets & $tW$ & $WW$ & $WZ$ & $t\bar{t}W$&$W4b$ \\
		\midrule
		BP1 & 0.394 & 0.454 & 4.890 & 0.003 & 0.226 & 0.248 & 0.000 & 0.164 \\
		BP2 & 0.760 & 1.362 & 9.781 & 0.005 & 0.226 & 0.551 & 0.002 & 0.382 \\
		BP3 & 0.721 & 1.362 & 9.781 & 0.005 & 0.226 & 0.579 & 0.002 & 0.376 \\
		BP4 & 0.685 & 1.362 & 9.781 & 0.005 & 0.226 & 0.551 & 0.002 & 0.377 \\
		BP5 & 0.782 & 5.675 & 22.007 & 0.007 & 0.339 & 7.830 & 0.008 & 0.885 \\
		BP6 & 0.694 & 2.724 & 17.117 & 0.006 & 0.226 & 2.371 & 0.007 & 0.641 \\
		BP7 & 0.611 & 2.724 & 17.117 & 0.006 & 0.226 & 2.426 & 0.007 & 0.639 \\
		BP8 & 0.663 & 12.713 & 30.565 & 0.009 & 0.452 & 23.242 & 0.015 & 1.264 \\
		BP9 & 0.532 & 18.161 & 37.901 & 0.011 & 0.452 & 26.467 & 0.028 & 1.556 \\
		\bottomrule
	\end{tabular}
	\caption{Signal (BPs for $W^+h_2$) and background cross-sections at $\sqrt{s}=13.6$ \TeV after selection cuts in Table~\ref{tab:cuts}. All values are in femtobarn (\fb).}
	\label{tab:xs_after_cuts}
\end{table}

\begin{table}[htb!]
	\centering
	\setlength{\tabcolsep}{4pt}
	\renewcommand{\arraystretch}{1.2}
	\begin{tabular}{lcccccccc}
		\toprule
		\multirow{2}{*}{BP} & \multicolumn{8}{c}{Cross-section [\fb] after selection cuts} \\
		\cmidrule(lr){2-9}
		&$Xs(BP_i)$& $t\bar{t}$+jets & $W$+jets & $tW$ & $WW$ & $WZ$ & $t\bar{t}W$&$W4b$ \\
		\midrule
		BP1 & 0.304 & 0.681 & 5.836 & 0.003 & 0.228 & 0.175 & 0.000 & 0.114 \\
		BP2 & 0.532 & 1.135 & 10.837 & 0.007 & 0.228 & 0.315 & 0.002 & 0.257 \\
		BP3 & 0.511 & 1.135 & 10.837 & 0.007 & 0.228 & 0.315 & 0.001 & 0.257 \\
		BP4 & 0.497 & 1.135 & 11.671 & 0.007 & 0.228 & 0.385 & 0.002 & 0.259 \\
		BP5 & 0.556 & 7.039 & 25.009 & 0.010 & 0.683 & 7.739 & 0.006 & 0.623 \\
		BP6 & 0.391 & 1.135 & 11.671 & 0.007 & 0.228 & 0.385 & 0.002 & 0.247 \\
		BP7 & 0.321 & 1.589 & 11.671 & 0.007 & 0.000 & 0.420 & 0.002 & 0.309 \\
		BP8 & 0.467 & 11.807 & 29.178 & 0.013 & 0.683 & 16.950 & 0.014 & 0.941 \\
		BP9 & 0.346 & 17.256 & 30.011 & 0.013 & 0.683 & 17.720 & 0.018 & 1.066 \\
		\bottomrule
	\end{tabular}
	\caption{Signal (BPs for $W^-h_2$) and background cross-sections at $\sqrt{s}=13.6$ \TeV after selection cuts in Table~\ref{tab:cuts}. All values are in femtobarn (\fb).}
	\label{tab:xs_after_cuts-wm}
\end{table}

\section{$K$-factors}
\setcounter{table}{0}
\label{sec:K-Factor}
The LO cross sections for both the signal and background processes can be rescaled using the corresponding $K$-factors to account for higher-order QCD corrections. The $K$-factors adopted for the signal and background processes are summarized in Tables~\ref{tab:without_decay} and ~\ref{tab:kfactors}, respectively. The $K$-factors for the signal are obtained using the official recommendations of the LHC Higgs Cross Section Working Group~\cite{13p6numbers}. K-factors for the signal are obtained using the official recommendations from the Higgs Working Group \cite{13p6numbers} for the production rate of an SM-like scalar of that mass rescaled by $\kappa_2^2$ and divided by the leading order cross section obtained from \texttt{MadGraph\_aMC@NLO}. 

\renewcommand{\arraystretch}{1.1}
\setlength{\tabcolsep}{0.06cm}
\begin{table}[H]
    \centering
    \begin{tabular}{|c|ccc|cccc|ccc|}
        \hline
        \multirow{2}{*}{BP} & \multicolumn{3}{c|}{Masses [GeV]} & \multicolumn{4}{c|}{Cross-sections $\sigma$ [\pb]} & \multicolumn{3}{c|}{$K$-Factors} \\
        \cline{2-11}
        & $M_1$ & $M_2$ & $M_2-2M_1$ & $W^+h_2$ & $W^-h_2$& $qq(Zh_2)$ & $gg(Zh_2)$ & $KW^+$ & $KW^-$&$KZ$ \\
        \hline
BP1  & 20.13 & 42.96 & 2.70  &  0.5425 (7)  &  0.3805 (5)  &  0.4421 (5)  &   0.00666 (3) & 1.50 & 1.50 & 1.63 \\
BP2  & 24.84 & 55.25 & 5.56  &  0.3419 (4)  &  0.2364 (3)  &  0.2839 (3)  &  0.00616  (2) & 1.31 & 1.30 & 1.39 \\
BP3  & 21.60 & 58.31 & 15.12 &  0.3105 (4)  &  0.2139 (3)  &  0.2589 (3)  &  0.00612  (2) & 1.17 & 1.16 & 1.25 \\
BP4  & 30.16 & 70.16 & 9.83  &  0.2012 (2)  &  0.1368 (2)  &  0.1698 (2)  &  0.00556  (2) & 1.23 & 1.22 & 1.32 \\
BP5  & 22.27 & 79.57 & 35.03 &  0.1485 (2)  &  0.0999 (1)  &  0.1264 (1)  &  0.00521  (2) & 1.17 & 1.16 & 1.28 \\
BP6  & 36.70 & 79.95 & 6.54  &  0.1448 (2)  &  0.0974 (1)  &  0.1233 (1)  &  0.00514  (2) & 1.20 & 1.19 & 1.31 \\
BP7  & 37.83 & 90.12 & 14.46 &  0.1077 (1)  &  0.07171(8)  & 0.0924 (1)  &  0.00482  (2)  & 1.16 & 1.16 & 1.29 \\
BP8  & 20.92 & 98.23 & 56.38 &  0.08575 (9) & 0.05663 (6)  & 0.07384 (9)  & 0.00411  (2)  & 1.08 & 1.07 & 1.22 \\
BP9  & 20.38 & 118.72& 77.57 &   0.05142 (6)& 0.03330 (4)  & 0.04455 (5)  &  0.003777 (5) & 1.04 & 1.03&  1.21 \\
        \hline
    \end{tabular}
    \caption{Leading-order cross-sections (in \pb) at the LHC with $\sqrt{s}=13.6$ \TeV, using Eqs.~(\ref{based_cuts1}) and (\ref{based_cuts2}). The signals correspond to the production channel ($W^+h_2$), ($W^-h_2$), and ($qq$ and $gg\to Zh_2$).}
    \label{tab:without_decay}
\end{table}

\begin{table}[H]
    \centering
    \setlength{\tabcolsep}{4pt}
    \renewcommand{\arraystretch}{1.3}
    \begin{tabular}{lcc}
        \toprule
        \textbf{Process} & \textbf{LO $\to$ NLO} & \textbf{NLO $\to$ NNLO} \\
        \midrule
        $t\bar{t}$ + jets \cite{Kidonakis:2023juy} & $1.46$ & $1.1$ \\
        & $1.40 - 1.59$ (0 jet) & $1.19$ (0 jet) \\
        & $0.97 - 1.29$ (1 jet) & $1.10$ (1 jet) \\
        \midrule
        $W$ + 1 jet (\cite{Huston:2010xp}) & $1.21 - 1.32$ & $1.42$ \\
        \midrule
        $W$ + 2 jets (\cite{Huston:2010xp}) & $0.89 - 0.88$ & $1.10$ \\
        \midrule
        $Z$ + jets~\cite{Boughezal:2015ded, Campbell:2002tg} & & \\
        & $1.2 - 1.3$ ($Z$ + 1 jet) & $1.0 - 1.2$ ($Z$ + 2 jets) \\
        \midrule
        $tW$ (\cite{Giammanco:2017xyn, Berger:2016oht})&1.35 - 1.40& 1.0 \\
        \midrule
        $ttW$ (\cite{Frederix:2017wme}) & $1.6$ & -- \\
        \midrule
        $WW$ (\cite{Grazzini:2016ctr}) & $1.30 - 1.40$ & $1.10 - 1.15$ \\
        \midrule
        $ZZ$ \cite{Cascioli:2014yka}, \cite{Buonocore:2021fnj} & $1.35 - 1.45$ & $1.50 - 1.65$ (global K-factor) \\
        \midrule
        $WZ$ (\cite{Grazzini:2017ckn}) & $1.30 - 1.40$ & $1.08 - 1.12$ \\
        \midrule
        $t\bar{t}Z$ \cite{Kidonakis:2024lht} & & \\
        & 1.32 & 1.09 \\
        \midrule
        \bottomrule
    \end{tabular}
    \caption{K-factors for background processes. Ranges indicate scale variations or jet multiplicities.}
    \label{tab:kfactors}
\end{table}
Taking into account the ranges of $K$-factors given above in Table~\ref{tab:kfactors}, the updated significances are presented in Table~\ref{tab:significance_Updated}. We emphasize that no $K$-factors are applied to the $t\bar{t}+$jets and $W+$jets backgrounds, since these samples are generated using matching and merging.
\begin{table}[htpb!]
	\centering
	\setlength{\tabcolsep}{4pt}
	\renewcommand{\arraystretch}{1.2}
	\begin{tabular}{lccccccccccc}
		\toprule
		\multirow{2}{*}{$\mathcal{L}$ [\fb$^{-1}$]} & \multirow{2}{*}{Channel} & \multicolumn{9}{c}{Benchmark Points} \\
		\cmidrule(lr){3-11}
		& & BP1 & BP2 & BP3 & BP4 & BP5 & BP6 & BP7 & BP8 & BP9 \\
		\midrule
		\multirow{3}{*}{300} & 
        $W^+h_2$ & 4.071 & 4.806 & 4.075 & 4.076 & 2.521 & 2.935 & 2.496 & 1.426 & 0.994 \\
		& $W^-h_2$ & 2.921 & 3.300 & 2.832 & 2.799 & 1.687 & 2.153 & 1.708 & 1.073 & 0.725 \\
		& Combined & 4.912 & 5.726 & 4.879 & 4.839 & 2.959 & 3.639 & 3.021 & 1.775 & 1.226 \\
		\midrule
		\multirow{3}{*}{3000} & 
        $W^+h_2$ & 12.874 & 15.199 & 12.888 & 12.890 & 7.972 & 9.281 & 7.894 & 4.510 & 3.145 \\
		& $W^-h_2$ & 9.238 & 10.436 & 8.956 & 8.852 & 5.334 & 6.807 & 5.403 & 3.393 & 2.294 \\
		& Combined & 15.532 & 18.106 & 15.430 & 15.303 & 9.357 & 11.508 & 9.552 & 5.614 & 3.876 \\
		\bottomrule
	\end{tabular}
\caption{Updated discovery significances after applying the corresponding $K$-factors to the signal and background processes for the $W^{+}h_2$, $W^{-}h_2$, and combined $W^{+}h_2+W^{-}h_2$ channels at $\sqrt{s}=13.6$~\TeV, assuming integrated luminosities of $\mathcal{L}=300$ and $3000$~\fb$^{-1}$.}
\label{tab:significance_Updated}
\end{table}

After including the corresponding $K$-factors for both the signal and background processes, the discovery significance generally increases for BP1--BP7. The largest enhancement is obtained for BP1, for which the combined significance increases from $3.48$ to $4.91$ at $\mathcal{L}=300~\fb^{-1}$, corresponding to an improvement of approximately $41\%$. For BP2, the combined significance increases from $4.61$ to $5.73$, thereby reaching the discovery threshold. Moderate improvements of approximately $11\%$--$17\%$ are obtained for BP3--BP7. In contrast, the impact is small for BP8, while the significance for BP9 decreases slightly. This behavior follows from the relatively small signal $K$-factors for the heavier benchmark points, combined with the larger higher-order corrections applied to some background processes.
To quantify the impact of the $K$-factors, we define the absolute and relative changes in the significance as
\begin{equation}
    \Delta Z = Z_{K}-Z_{\mathrm{LO}},
    \qquad
    \delta_Z =
    \frac{Z_{K}-Z_{\mathrm{LO}}}{Z_{\mathrm{LO}}}
    \times 100\%,
    \label{eq:significance_variation}
\end{equation}
where $Z_{\mathrm{LO}}$ and $Z_K$ denote the significances obtained before and
after applying the $K$-factors, respectively.

\subsection*{Comparison of the combined significances}

Table~\ref{tab:combined_significance_comparison} compares the combined
$W^+h_2+W^-h_2$ significances before and after applying the $K$-factors at
$\mathcal{L}=300~\fb^{-1}$. The corresponding absolute and relative changes
are also reported.

\begin{table}[htpb!]
    \centering
    \setlength{\tabcolsep}{6pt}
    \renewcommand{\arraystretch}{1.2}
    \begin{tabular}{lcccc}
        \toprule
        BP
        & $Z_{\mathrm{LO}}$
        & $Z_K$
        & $\Delta Z$
        & $\delta_Z$ [\%] \\
        \midrule
        BP1 & 3.480 & 4.912 & $+1.432$ & $+41.1$ \\
        BP2 & 4.610 & 5.726 & $+1.116$ & $+24.2$ \\
        BP3 & 4.400 & 4.879 & $+0.479$ & $+10.9$ \\
        BP4 & 4.150 & 4.839 & $+0.689$ & $+16.6$ \\
        BP5 & 2.660 & 2.959 & $+0.299$ & $+11.2$ \\
        BP6 & 3.160 & 3.639 & $+0.479$ & $+15.2$ \\
        BP7 & 2.700 & 3.021 & $+0.321$ & $+11.9$ \\
        BP8 & 1.750 & 1.775 & $+0.025$ & $+1.4$  \\
        BP9 & 1.250 & 1.226 & $-0.024$ & $-1.9$  \\
        \bottomrule
    \end{tabular}
    \caption{Comparison of the combined $W^+h_2+W^-h_2$ discovery
    significances before and after applying the signal and background
    $K$-factors at $\sqrt{s}=13.6~\TeV$ and
    $\mathcal{L}=300~\fb^{-1}$. The absolute and relative changes are
    defined in Eq.~\eqref{eq:significance_variation}.}
    \label{tab:combined_significance_comparison}
\end{table}

Since no systematic uncertainties are included, the Asimov significance scales
with the square root of the integrated luminosity,
\begin{equation}
    Z_A \propto \sqrt{\mathcal{L}}.
    \label{eq:significance_luminosity_scaling}
\end{equation}
Consequently, the relative changes at $\mathcal{L}=3000~\fb^{-1}$ are
effectively identical to those obtained at $\mathcal{L}=300~\fb^{-1}$, up to
rounding effects.

\subsection*{Comparison of the individual and combined channels}

The relative changes in the significances of the $W^+h_2$, $W^-h_2$, and
combined channels are summarized in
Table~\ref{tab:relative_significance_comparison}.

\begin{table}[htpb!]
    \centering
    \setlength{\tabcolsep}{7pt}
    \renewcommand{\arraystretch}{1.2}
    \begin{tabular}{lccc}
        \toprule
        BP
        & $\delta_Z(W^+h_2)$ [\%]
        & $\delta_Z(W^-h_2)$ [\%]
        & $\delta_Z(\mathrm{Combined})$ [\%] \\
        \midrule
        BP1 & $+39.4$ & $+42.5$ & $+41.1$ \\
        BP2 & $+23.2$ & $+25.0$ & $+24.2$ \\
        BP3 & $+10.1$ & $+11.5$ & $+10.9$ \\
        BP4 & $+15.8$ & $+17.1$ & $+16.6$ \\
        BP5 & $+11.5$ & $+11.0$ & $+11.2$ \\
        BP6 & $+15.6$ & $+14.5$ & $+15.2$ \\
        BP7 & $+11.9$ & $+12.4$ & $+11.9$ \\
        BP8 & $+1.1$  & $+1.2$  & $+1.4$  \\
        BP9 & $-1.6$  & $-2.0$  & $-1.9$  \\
        \bottomrule
    \end{tabular}
    \caption{Relative changes in the discovery significances after applying
    the corresponding signal and background $K$-factors for the
    $W^+h_2$, $W^-h_2$, and combined channels at
    $\mathcal{L}=300~\fb^{-1}$.}
    \label{tab:relative_significance_comparison}
\end{table}

\subsection*{Physics interpretation}

The largest enhancement is obtained for BP1, for which the signal
$K$-factors are
\begin{equation}
    K_S^{W^+h_2}=K_S^{W^-h_2}=1.50.
\end{equation}
These corrections are substantially larger than the effective correction to
the total background. Consequently, the combined significance increases from
\begin{equation}
    Z_{\mathrm{LO}}=3.48
    \qquad\text{to}\qquad
    Z_K=4.912,
\end{equation}
corresponding to an improvement of approximately $41\%$ at
$\mathcal{L}=300~\fb^{-1}$.

For BP2, the combined significance increases from
\begin{equation}
    4.61 \longrightarrow 5.726,
\end{equation}
thereby exceeding the conventional $5\sigma$ discovery threshold. For
BP3--BP7, the signal $K$-factors lie approximately between $1.16$ and $1.23$.
This leads to moderate improvements in the combined significance, ranging from
approximately $11\%$ to $17\%$.

For BP8, the signal corrections are relatively small,
\begin{equation}
    K_S^{W^+h_2}=1.08,
    \qquad
    K_S^{W^-h_2}=1.07.
\end{equation}
The increase in the signal yield is therefore almost compensated by the
higher-order corrections applied to the background processes. As a result, the
combined significance increases by only approximately $1.4\%$.

For BP9, the signal $K$-factors are even closer to unity,
\begin{equation}
    K_S^{W^+h_2}=1.04,
    \qquad
    K_S^{W^-h_2}=1.03.
\end{equation}
At the same time, several background processes, including $tW$, $WW$, $WZ$,
and $t\bar{t}W$, receive larger corrections. The total background yield
therefore increases proportionally more than the signal yield, leading to a
small reduction in the combined significance,
\begin{equation}
    1.250 \longrightarrow 1.226,
\end{equation}
which corresponds to a relative decrease of approximately $1.9\%$.

\subsection*{Impact on the expected sensitivity}

At $\mathcal{L}=300~\fb^{-1}$, the inclusion of the $K$-factors changes the
expected sensitivity as follows:
\begin{itemize}
    \item BP2 increases from $4.61\sigma$ to $5.726\sigma$ and therefore
    reaches the discovery level.

    \item BP1 increases substantially from $3.48\sigma$ to $4.912\sigma$,
    approaching the $5\sigma$ discovery threshold.

    \item BP3 and BP4 reach combined significances of $4.879\sigma$ and
    $4.839\sigma$, respectively, and therefore remain close to the discovery
    threshold.

    \item BP7 increases from $2.70\sigma$ to $3.021\sigma$ and reaches the
    evidence level.

    \item BP5 remains marginally below the evidence threshold, with
    $Z_K=2.959\sigma$.

    \item BP8 and BP9 remain below the evidence threshold.
\end{itemize}

At $\mathcal{L}=3000~\fb^{-1}$, BP1--BP8 reach combined significances above
$5\,\sigma$. In particular, the BP8 significance increases slightly from
$5.54\,\sigma$ to $5.61\,\sigma$. In contrast, BP9 remains below the discovery
threshold, and its significance decreases slightly from $3.96\,\sigma$ to
$3.88\,\sigma$ after applying the $K$-factors.

\bibliographystyle{JHEP}
\bibliography{refference}

\end{document}